\documentclass[12pt]{article}

\usepackage{newtxtext,newtxmath}

\usepackage{graphicx}

\usepackage[letterpaper,margin=1in]{geometry}

\renewenvironment{abstract}
	{\quotation}
	{\endquotation}

\date{}

\makeatletter
\renewcommand{\fnum@figure}{\textbf{Figure \thefigure}}
\renewcommand{\fnum@table}{\textbf{Table \thetable}}
\makeatother

\usepackage{scicite}

\usepackage{siunitx}

\usepackage{url}

\newcommand{\bra}[1]{\langle #1 |}
\newcommand{\ket}[1]{|#1\rangle}
\newcommand{\abs}[1]{|#1|}
\newcommand{\expval}[1]{\langle #1 \rangle}

\def\scititle{
	Decoherence of Majorana Qubits by 1/f Noise
}
\title{\bfseries \boldmath \scititle}

\author{
	Authors:
    A.~Alase$^{1,2}$,
	M.~C.~Goffage$^{3}$,
    M.~C.~Cassidy$^{3}$,
	S.~N.~Coppersmith$^{3,\ast}$\and
	\small$^{1}$School of Physics and Centre for Engineered Quantum Systems, University of Sydney, Sydney, NSW 2006, Australia\and
	\small$^{2}$Department of Physics, Concordia University, Montreal, QC H4B 1R6, Canada\\
    \small$^{3}$School of Physics, University of New South Wales, Sydney, NSW 2052, Australia\and
	\small$^\ast$Corresponding author. Email: s.coppersmith@unsw.edu.au\and
}

\begin{document} 

\maketitle

\begin{abstract} 
\bfseries \boldmath
\centering{{\bf Abstract}}
\\
\flushleft
~~~~~~Qubits based on Majorana zero modes (MZMs) in superconductor–
semiconductor nanowires have attracted intense interest due to claims that their error rates are suppressed exponentially with increasing nanowire length or decreasing temperature. 
However, here we show that these qubits are subject to substantial decoherence resulting from the high-frequency components of 1/f charge noise, which is ubiquitous in the materials surrounding the nanowire. 
This process excites quasiparticles in the bulk of the topological superconductor that cause qubit decoherence even under otherwise ideal conditions. 
Increasing nanowire capacitance suppresses this mechanism but exposes the qubits to decoherence from externally-generated quasiparticles. Therefore, achieving high-fidelity MZM qubits using superconductor–
semiconductor nanowires will require engineering strategies and compromises very similar to those needed for conventional superconducting qubits.\
\end{abstract}
\hfil\\
\noindent
{\bf Teaser: A process that seriously limits the quantum coherence of Majorana qubits hosted in InAs nanowires is identified and characterized.} 

\noindent

{\noindent \bf Introduction.}
The goal of realizing and manipulating qubits encoded in
Majorana Zero Modes (MZMs) in nanowires that support topologically-nontrivial superconductivity \cite{Kitaev_2001_Physics_uspekhi_44, Lutchyn_2010_PRL_105_01, Oreg_2010_PRL_17_02, Aghaee_2023_PRB_107_23, Alicea_2012_IOP_75, Sarma_2015_npj_01}
has drawn enormous interest and investment over the past decade~\cite{Aghaee_2023_PRB_107_23, Mourik_2012_Science_336_07, Albrecht_2016_Nature_531_09, aghaee_2025_nature_638_55, Aasen2025arxiv.2502.12252}. These qubits are predicted to be far less prone to decoherence than alternative architectures~\cite{bravyi_2022_JAppPhys_132,Burkard_2023_RevModPhys_95,knapp2018dephasing}, and therefore much more advantageous for scaling up to large processors that could yield tremendous computational advantage.

 MZM qubits have been believed to be extremely resistant to decoherence because they are composed of degrees of freedom in a superconducting condensate that are spatially delocalized, which plausibly provides protection from localized decoherence processes affecting other qubits. However, it is known that quasiparticles, which are unpaired electrons mobile in the superconducting condensate, can be extremely detrimental to MZM qubits which encode quantum information in the joint parity of MZM pairs. Quasiparticles that reach the end of the wire~\cite{karzig_2021_phys_rev_let_126} can interact strongly with the MZMs, thus changing the joint MZM parity 
 and resulting in qubit errors~\cite{knapp_2018_Quantum_88,Alase_2024_Phys_Rev_Res_6}, in a process known as quasiparticle poisoning~\cite{Rainis_2012_PRB_17, Karzig_2017_Phys_Rev_B_95}.
 These quasiparticles may come from sources extrinsic to the qubit, for example from coupling to normal metal leads, 
or they can be generated intrinsically through thermal excitations or interaction with high frequency radiation that breaks the Cooper pairs~\cite{pedrocchi15, knapp_2018_PRB_97_12, karzig_2021_phys_rev_let_126}. Up until now, it has been believed that errors due to 
intrinsic quasiparticle poisoning can be made negligibly small at cryogenic temperatures if the experiment is shielded from external high-frequency radiation because intrinsic quasiparticles are suppressed exponentially by lowering the temperature%
~\cite{knapp_2018_PRB_97_12, karzig_2021_phys_rev_let_126, aghaee_2025_nature_638_55}.

Here we show that small fluctuations in chemical potential resulting from two level fluctuators (TLFs), which are ubiquitous in the materials making up these qubits~\cite{Dutta:1981p497,paladino20141}, excite quasiparticle pairs in the bulk of the nanowire.
While the excitation of quasiparticle pairs is suppressed exponentially in the asymptotic limit of zero temperature,
the methods used in Ref.~\cite{aghaee_2025_nature_638_55}
are unlikely to achieve the asymptotic regime.
Using realistic parameters from recent InAs-based devices~\cite{Aghaee_2023_PRB_107_23,aghaee_2025_nature_638_55,Aghaee.arxiv.2507.08795} and the Microsoft roadmap~\cite{Aasen2025arxiv.2502.12252}, we show that this noise would cause a MZM qubit to decohere on a time scale of microseconds. This decoherence mechanism can be mitigated by increasing the capacitance of the qubits, but at the cost of removing the protection from decoherence induced by externally generated quasiparticles.
\textcolor{black}{Therefore, 
Microsoft's current architecture and materials platform will require engineering tradeoffs that are very similar to those needed for non-topological superconducting qubits to achieve high-fidelity qubits.}

\begin{figure}[t]
    \centering
\includegraphics[width=0.9\linewidth]{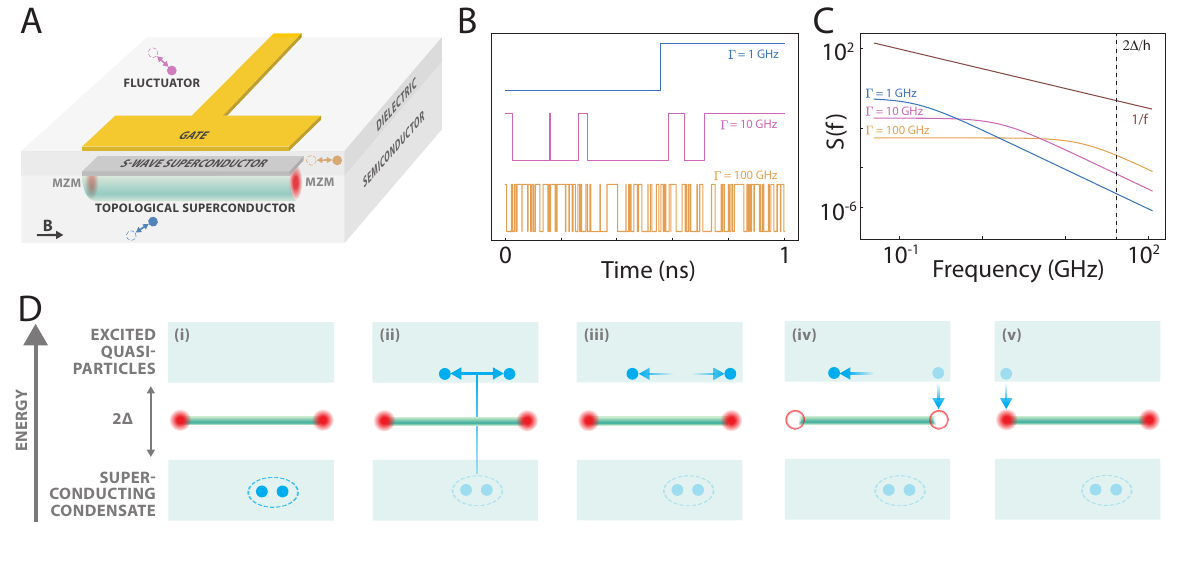} 
    \caption{\small \bf Quasiparticle generation in MZM nanowires by TLFs. \rm
  {A}: Schematic of a superconducting-semiconducting nanowire device hosting MZMs. Atomic scale defects found in the materials surrounding the nanowire give rise to two-level fluctuators (TLFs) with different transition frequencies. {B}: Each TLF undergoes a series of sudden transitions between its states, each causing an instantaneous step in the chemical potential in the nanowire. {C}:  The resulting frequency spectrum has components extending above the superconducting gap $\Delta$. A 1/f noise spectrum arises from an ensemble of two-level fluctuators (TLFs)~\cite{Dutta:1981p497}. {D}: \textcolor{black}{Time sequence of dephasing by excited quasiparticle pairs:} (i) At zero temperature, bound Cooper pairs in the superconducting condensate are separated in energy from excited quasiparticles by twice the superconducting gap. (ii) Small sudden changes in chemical potential excite quasiparticle pairs from the superconducting condensate, (iii) which then travel to the ends of the wire and (iv) interact with the MZMs, changing the parity or (v) causing a qubit error.    \label{fig:calculation_plan}
}
\end{figure}

{\noindent \bf Decoherence mechanism.}
Figure~\ref{fig:calculation_plan} illustrates a MZM nanowire device and the mechanism for bulk quasiparticle generation from TLFs. A one-dimensional nanowire is formed within a semiconducting substrate by depleting the regions around a narrow superconducting strip with an external gate, as shown in Figure~\ref{fig:calculation_plan}A. MZMs are formed at each end of the nanowire by adjusting the external magnetic field B and chemical potential $\mu$ so that the nanowire is tuned into a topologically non-trivial phase. Two level fluctuators (TLFs) are present in the materials surrounding the nanowire~\cite{Dutta:1981p497,doi:10.1126/science.abb2823}. 
These atomic-scale defects behave as 
two-level systems described by characteristic switching rates $\Gamma$, that still exhibit transitions close to zero temperature. The combined effects of these many sudden jumps (Figure~\ref{fig:calculation_plan}B) results in a frequency spectrum $S(f) \propto~1/f$ that, crucially, has components that extend beyond the topological superconducting gap $\Delta$ (Figure~\ref{fig:calculation_plan}C)~\cite{paladino20141,Covington2000p5192}. Charge noise from TLFs is a known phenomenon affecting superconducting and semiconducting qubits~\cite{doi:10.1126/science.abb2823}. 
Direct observations of charge noise following a $1/f$ spectrum up to 10~{\rm MHz} at lower temperatures have been performed in semiconductor qubits~\cite{connors_charge-noise_2022}, and 
evidence of out of equilibrium TLFs with an effective temperature of up to several hundred millikelvin
that can be present in superconducting and semiconducting devices \cite{Lucas2023QuantumBath, Huang:2024p772,PhysRevApplied.16.014019, Muller_2019, astafiev2004quantum}.

Bound Cooper pairs in the superconducting condensate are separated in energy from excited quasiparticles by twice the superconducting gap (Figure~\ref{fig:calculation_plan}D). A TLF switching between its two states causes small, sudden jumps in the nanowire chemical potential $\mu$ that are far smaller than what is required to bring the device out of the topological regime. 
However, when the chemical potential $\mu$ jumps slightly and suddenly from $\mu_1$ to $\mu_2$, the ground and excited quasiparticle states also undergo a change. The original ground state at chemical potential $\mu_1$, although unaffected by the sudden change, can now be thought of as a superposition of the ground state and excited quasiparticle states of the instantaneous Hamiltonian with chemical potential $\mu_2$. Importantly, this quantum superposition of the ground state and excited states dephases quickly, so that successive changes of $\mu$ will excite more quasiparticles.

\textcolor{black}{This quasiparticle pair excitation process is suppressed if the TLFs are at low enough temperatures, but we will show that the topological gaps measured in Ref.~\cite{aghaee_2025_nature_638_55,Aghaee.arxiv.2507.08795} are not high enough to eliminate this decoherence source at the TLF temperatures typical of semiconducting qubit architectures~\cite{Huang:2024p772}.}

\textcolor{black}{In a defect-free nanowire the quasiparticles in the excited pairs have opposite momenta, so they will travel to opposite ends of the host nanowire, where, as shown in Ref.~\cite{karzig_2021_phys_rev_let_126}, each quasiparticle is absorbed with high probability by an MZM,
which results
in a qubit error~\cite{knapp_2018_PRB_97_12}.} 
If the nanowires have defects so that quasiparticle motion is diffusive, with probability 1/3 the quasiparticles in the pair diffuse to opposite ends of the nanowire~\cite{goffage2025leakage} and yield a qubit error.
There is a temporary parity error in the time interval during which one but not both of the quasiparticles has been absorbed,  but the rate of quasiparticle absorption is much higher than the rate of creation, so measurements of the parity will not exhibit a decay~(see Supplementary Sec.~\ref{supp_sec:diffusion}).
The spatial separation between the MZMs responsible for topological protection does not protect against this decoherence mechanism, which occurs even if the nanowires themselves have no imperfections~\cite{goffage2025leakage}.

{\noindent \bf Results.}
To estimate the rate of qubit errors, we first calculate the bulk contribution to the probability of excitation of a pair of quasiparticles, $P_{\rm QPP}^{(1)}$, from one sudden change of $\mu$.  
\textcolor{black}{We use the result that the low energy behavior of Microsoft's nanowire system tuned to the topological regime is captured by a Kitaev Hamiltonian of spinless fermions with p-wave superconducting pairing
with Kitaev chain parameters that can be derived from the properties of the nanowire system~\cite{Alicea_2012_IOP_75,Aghaee_2023_PRB_107_23}.}
We consider the Kitaev Hamiltonian 
with $N$ sites with a circular geometry tuned in the topological superconducting state~\cite{Kitaev_2001_Physics_uspekhi_44}:
\begin{equation}
   H_K = \sum_{i=1}^N \left [ -\mu c_i^\dagger c_i - \frac{1}{2} \left (w c_i^\dagger c_{i+1} + \Delta c_i c_{i+1} + \rm{H.c.} \right )
   \right ]~.
   \label{eq:Kitaev}
\end{equation}
Here, $\mu$ is the chemical potential, $w$ is the nearest-neighbor hopping strength of the fermions, $\Delta$ is the superconducting gap, $c_i^\dagger$ ($c_i$) are operators that create (annihilate) a spinless fermion at site $i$, and H.c.\ denotes the Hermitian conjugate. 
We perform a spatial Fourier transform and calculate the probability $P_{\rm QPP}^{(1)}(k)$ of excitation of a quasiparticle pair with wavevectors $(k,-k)$, and then sum over all such pairs to obtain $P_{\rm QPP}^{(1)}$. In the Methods, we discuss
how $P_{\rm QPP}^{(1)}(k)$ relates to the overlap of the eigenvectors of
$H_k$ for the values of $\mu$ before and after the jump.
For a small chemical potential change $\delta \mu$, we can express $P_{\rm QPP}^{(1)}(k)$ 
in terms of 
$\delta \mu$, the Fermi velocity $v_F = wa/\hbar$, where $a$ is the lattice constant of the chain, the nanowire length $\mathscr{L}$, and the topological superconducting gap $\Delta$. 
The total probability of exciting a quasiparticle pair by a single jump $\delta \mu$ of the chemical potential in a nanowire of length $\mathscr{L}$ is found by summing over all $k$ values, yielding
 \begin{equation}
     P_{\rm QPP}^{(1)} =
          \frac{\mathscr{L}}{16}
           \frac{(\delta\mu)^2}{\hbar v_F \Delta}~.
           \label{eq:PTOT_in_main_text}
 \end{equation}

For a single TLF with switching rate $\Gamma$, if $\Gamma$ is sufficiently slow that the state dephases between switches, then the probability of excitation of quasiparticle pairs will simply add up with successive transitions of the TLF, yielding the rate of quasiparticle pair excitation due to a single TLF to be $R_{\rm QPP} = \Gamma P_{\rm QPP}^{(1)}$. For switching rates comparable to or higher than $\Delta/h$, the dephasing is not perfect, and so the rate of quasiparticle pair excitation is instead found to be
\begin{equation}
\label{eq:R_QPP_with_F_included}
R_{\rm QPP}(\Gamma) = \mathscr{F}\Gamma P_{\rm QPP}^{(1)} = \mathscr{F}\frac{\mathscr{L}}{16}
           \frac{\Gamma(\delta\mu)^2}{\hbar v_F \Delta}~.
\end{equation}
with $\mathscr{F} \in [0,1]$ a multiplicative factor that is $1$ when $\Gamma$ is very small and $0$ when $\Gamma$ is very large.
This expectation for $\mathscr{F}$ is confirmed by numerical simulations of the time evolution of the Kitaev Hamiltonian, Eq.~\ref{eq:Kitaev}, where the chemical potential fluctuates due to a single TLF and the probability of quasiparticle pair generation is calculated.  This model has been studied previously for low-frequency noise with Gaussian time correlations (work that did not reveal significant quasiparticle excitation because of the lack of high-frequency components in the noise)
~\cite{Mishmash_2020_PRB_101}, as well as for a ramp function~\cite{goffage2025leakage}.
Our simulations, described in detail in Supplementary Section~\ref{supp_sec:numerics_details}, 
demonstrate that $R_{\rm QPP}(\Gamma)$ is largest when $\Gamma$ is of order $4\Delta/\hbar$, as demonstrated in Fig.~\ref{fig:F2}B.
Because the TLF transition rates are distributed logarithmically in frequency~\cite{Dutta:1981p497}, the sum over TLFs is dominated by its largest term, so evaluating the excitation rate at the value of $\Gamma$ that maximizes the rate yields a reasonable estimate of the total quasiparticle pair excitation rate (see Supplementary Sec.~\ref{sec:Gamma_max_calculations} for details).

The decoherence rate in Eq.~\ref{eq:R_QPP_with_F_included} also depends on $\delta \mu$, the magnitude of the jumps of the chemical potential.
We find that $S_0$, the coefficient of $1/\omega$ in the noise spectrum, is $S_0 = (\delta\mu)^2/8$ (see Methods and Supplementary Sec.~\ref{supp_sec:delta_mu_derivation}), so that
\begin{equation}
    R_{\rm QPP} = \frac{\mathscr{F}\mathscr{L}S_0}{\pi\hbar^2v_{\rm F}}\frac{\Gamma}{\Delta}~.
    \label{eq:quasiparticle_excitation_rate}
\end{equation}
Ref.~\cite{aghaee_2025_nature_638_55} reports the value of $S_0^{\rm dot}$, where the power spectrum of the chemical potential fluctuations $S^{\rm dot}(\omega)$ is $S_0^{\rm dot}/\omega$, for a reference quantum dot with capacitance $C_{\rm dot} \approx 0.445$~fF located adjacent to a superconducting nanowire hosting MZMs.
Assuming the charge motions of the defects near a superconducting nanowire with capacitance $C_{\rm wire}$ are statistically similar to those near the reference dot, the power spectrum of the chemical potential fluctuations of the nanowire are $S(\omega)=S_0/\omega$, with~\cite{Yu.arxiv.2512.05644}
\begin{equation}
    \label{eq:noise_scaling_with_capacitance}
    S_0=S_0^{\rm dot}/(C_{\rm wire}/C_{\rm dot})^2~;
\end{equation}
this follows because a shift in charge $\delta q$ yields a voltage shift of $V=\delta q/C$, where $C$ is the relevant capacitance (see ~\cite{Yu.arxiv.2512.05644} and Supplementary Sec.~\ref{supp_sec:delta_mu_derivation} for details).
Requiring the capacitance of the wire to be small enough that the wire's charging energy $Q^2/2C$ is much greater than the electron temperature, needed to suppress poisoning from quasiparticles generated elsewhere in the system~\cite{aghaee_2025_nature_638_55,Aasen2025arxiv.2502.12252}, yields a maximum capacitance for the nanowire of $\approx 2.2~$fF (corresponding to a charging energy of about $36~\mu{\rm eV}$, which corresponds to $\sim 0.4$~K).
Supplementary Sec.~\ref{supp_sec:delta_mu_derivation} shows that using this value of the capacitance along with the measured noise spectrum and capacitance of the quantum dot~\cite{aghaee_2025_nature_638_55} yields $\delta\mu\approx 0.57~\mu$eV.

\textcolor{black}{So far our calculations have assumed that $T_{\rm eff}$, the  effective temperature of the TLFs, satisfies $k_BT_{\rm eff} \gg 2\Delta$, so that the TLFs possess enough energy to excite quasiparticles across the superconducting gap. In the opposite limit, where $2\Delta \gg k_BT_{\rm eff}$, quasiparticle pair creation is expected to be suppressed by the Boltzmann factor $\exp(-2\Delta/k_B T_{\rm eff})$. In practice, however, achieving low effective TLF temperatures is difficult in devices with complex multilayered gate structures, and where qubit operation and readout requires strong pulsed and RF electric fields that couple efficiently to the TLFs and drive them out of equilibrium. As a result, $T_{\rm eff}$ can substantially exceed both the electron temperature and the base temperature of the dilution refrigerator. For example, Ref.~\cite{Huang:2024p772} provides strong evidence for a TLF effective temperature of $\gtrsim$0.3~K in a silicon qubit platform, as discussed further in Supplementary Sec.~\ref{subsec:TLF_effective_temperature_data}. Under such conditions, $k_B T_{\rm eff}$  may become comparable, or even exceed $2\Delta$, implying that the exponential suppression of quasiparticle generation in  nanowires such as those studied in  Refs.~\cite{aghaee_2025_nature_638_55} may be weak.}


\textcolor{black}{To calculate the rate of quasiparticle pair excitation at finite $T_{\rm eff}$, we start with Eq.~\ref{eq:R_QPP_with_F_included}, approximate $\mathscr{F}$ as 
$\exp(-h\Gamma/(2\Delta))$, account for thermal effects via the exponential factor $\exp(-2\Delta/k_B T_{\rm eff})$,
and then find the maximum value of $R_{\rm QPP}(\Gamma)$ as a function of $\Gamma$ (see Eq.~\ref{eq:R_QPP_for_finite_temperature} in Methods).}
Our methods for determining relevant parameters such as the Fermi velocity $v_F$ (which depends on $\Delta$) from theory~\cite{Lutchyn_2010_PRL_105_01,Alicea_2012_IOP_75,Leijnse_2012,Aghaee_2023_PRB_107_23,Sarma_2015_npj_01} together with the experimental data in~\cite{aghaee_2025_nature_638_55} is described in Supplementary Sec.~\ref{supp_sec:nanowire_properties}, with parameter values given in Table~1.
We assume that there is a nearby TLF with a value of $\Gamma$ that yields a pair excitation rate close to the maximum; this assumption is reasonable because the nanowires used in Ref.~\cite{aghaee_2025_nature_638_55} have an area of about 60~nm~$\times$~3~$\mu m$, which is about 100 times larger than the area of Si/SiGe qubits that are affected by of order one TLF per two decades in frequency~\cite{connors_charge-noise_2022}.

Fig.~\ref{fig:F2}C shows the dependence of the quasiparticle pair excitation rate $R_{\rm QPP}$ on $\Delta$, the topological gap, for four different values of $T_{\rm eff}$.
The vertical lines are at $\Delta=16.8~\mu$eV and at $\Delta=23.8~\mu$eV, which are the median and the 90$^{\text{th}}$ percentile of topological gap values measured in Ref.~\cite{aghaee_2025_nature_638_55}
(see Supplementary Sec.~\ref{supp_sec:gap_histogram}).
Even at the lowest $T_{\rm eff}=50$~mK, the calculated rate of quasiparticle pair excitation $R_{\rm QPP}$ at the median gap of 16.8~$\mu$eV is about 3.8~kHz, and $R_{\rm QPP}$ grows quickly with $T_{\rm eff}$. 

We also note that the quasiparticle pair excitation rate is further enhanced if TLF transitions occur via activation over energy barriers. In semiconductors, the relevant attempt frequencies are typically in the THz range~\cite{Vineyard:1957p121}, implying that TLFs with transition rates on the order of $k_B T_{\rm eff}$ are associated with activation over substantial barriers. Consequently, the energy released during a TLF transition can be significantly larger than would be inferred from a naive estimate based solely on the effective TLF temperature. Our method for calculating the resulting quasiparticle pair excitation rate is described in the Methods (Eq.~\ref{eq:R_QPP_with_thermal_activation}). Fig.~\ref{fig:F2}D shows that the quasiparticle pair production rate, $R_{\rm QPP}$
for a TLF with transition rate $\Gamma=6.25$~GHz is substantially enhanced when the TLF transitions involve activation over a significant barrier. We note that there is experimental evidence that thermally activated TLFs are important at the lowest temperatures measured in SiGe quantum dot qubits~\cite{Ye:2024p235305}.

{\noindent \bf Decoherence of tetron qubits from excited quasiparticle pairs.}
We now discuss the implications of our results for the tetron qubits being developed by Microsoft Azure Quantum~\cite{Karzig_2017_Phys_Rev_B_95,Aghaee_2023_PRB_107_23,aghaee_2025_nature_638_55,Aghaee.arxiv.2507.08795}.
A tetron qubit is composed of two nanowires that host MZMs at their ends, along with a backbone composed of an s-wave superconductor, as illustrated in Fig.~\ref{fig:tetron}a~\cite{Karzig_2017_Phys_Rev_B_95}.
The tetron backbone ensures that the two nanowires have the same superconducting phase parameter and enables measurement of the qubit state via joint tunneling into both nanowires~\protect{\cite{Karzig_2017_Phys_Rev_B_95}}.
There is no electronic transport along the backbone between the two nanowires because its large superconducting gap far exceeds the topological gap, so the mechanisms leading to quasiparticle poisoning of the two nanowires are independent and hence the rate of quasiparticle poisoning in each nanowire is given by Eq.~\eqref{eq:quasiparticle_excitation_rate}. The two basis states of the tetron qubit are chosen to be the states where the zero-energy modes of the two nanowires are both empty or both occupied, respectively (see Figure~\ref{fig:tetron}A). The tetron qubit has two parity leakage states corresponding to only one of the two zero modes being occupied, and represent a form of decoherence of the qubit~\cite{Knapp2018modelingnoiseerror,Alase_2024_Phys_Rev_Res_6}.

Figure~\ref{fig:tetron}B illustrates the evolution of the tetron after a pair of quasiparticles has been excited in one of the nanowires. Because superconductivity pairs opposite momenta $k$ and $-k$, in a perfectly clean nanowire, the quasiparticles in the generated pair travel ballistically to opposite ends of the nanowire at the Fermi velocity. 
It is shown in Ref.~\cite{karzig_2021_phys_rev_let_126} that MZMs absorb quasiparticles efficiently, so the quasiparticles will be absorbed at opposite ends of the nanowire, which results in a $Z$ (dephasing) error~\cite{knapp_2018_PRB_97_12}.
There will be a short interval during which one quasiparticle but not the other has been absorbed by an MZM, during which the tetron is in a parity leakage state.
\textcolor{black}{The quasiparticle pair excitation process does not cause decay when Majorana parity is measured~\cite{aghaee_2025_nature_638_55,Aghaee.arxiv.2507.08795} because the time it takes both quasiparticles to be absorbed is much shorter than both the rate of quasiparticle pair excitation and the temporal resolution of the parity measurement (for details, see Supplementary Sec.~\ref{supp_sec:diffusion}).}

As the tetron consists of two nanowires, the dephasing rate $T_2^*$ of the tetron qubit is twice the rate of quasiparticle pair excitation in a single wire.  Thus, $T_2^*$ of a  tetron qubit made of two defect-free 3~$\mu$m nanowires, each with a quasiparticle pair production rate of $10^5$ pairs/s (appropriate for a gap of 25~$\mu$eV and $T_{\rm eff}=140$~mK) will have a $T_2^*$ of 5~$\mu$s.
This decoherence time is only slightly longer than the target measurement time on the Microsoft roadmap~\cite{Aasen2025arxiv.2502.12252} of 1~$\mu$s.
Disorder in the nanowire will scatter the quasiparticles so that they move diffusively, reducing the fraction of pairs absorbed by MZMs at opposite ends to 1/3 (see~\cite{goffage2025leakage} and Supplementary Sec.~\ref{supp_sec:diffusion}), 
but this modest reduction in poisoning rate is likely to be offset by other known deleterious effects of disorder on qubit properties~\cite{woods2021charge}.

We now address how to reduce the decoherence rate from quasiparticle pair excitation by 1/f charge noise.
Increasing the topological gap and decreasing the effective temperature could be effective in suppressing quasiparticle pair excitation.
However, achieving these improvements could be challenging because the measured topological gap is close to theoretical predictions~\cite{Alicea_2012_IOP_75,Aghaee_2023_PRB_107_23}, so that a change of the materials system is likely to be required. 
Improvements in materials that reduce the noise from TLFs will significantly improve coherence. Figure~\ref{fig:tetron}{\bf C} shows that the rate of quasiparticle pair excitation decreases as $1/S_0$, where $S_0$ is the coefficient quantifying the noise power.
Figure~\ref{fig:tetron}D shows the most promising avenue for improving qubit coherence, which is to lower the coefficient of the charge noise power $S_0$ by increasing
the relevant capacitance $C$.
The quasiparticle pair excitation rate is proportional to $1/C^2$ (see Eq.~\ref{eq:noise_scaling_with_capacitance}), and increasing qubit capacitance by more than two orders of magnitude is straightforward~\cite{Martinis2022p26}, so large increases in coherence time are feasible.
In fact, we expect the capacitance of tetron qubits to be much greater than that reported for the devices measured in Ref.~\cite{aghaee_2025_nature_638_55} because the structure includes a superconducting backbone that increases the capacitance~\cite{Karzig_2017_Phys_Rev_B_95,Aasen2025arxiv.2502.12252,Aghaee.arxiv.2507.08795}.
However, increasing the qubit capacitance also reduces the charging energy that protects the system against poisoning by quasiparticles generated elsewhere in the device. Strategies developed to suppress extrinsic quasiparticle poisoning in conventional superconducting qubits (see, e.g., Ref.~\cite{Pan2022p7196}) can also be applied to Majorana qubits. Nevertheless, such approaches are unlikely to provide coherence advantages beyond those already achieved in non-topological superconducting qubits. Given the slower gate speeds and more constrained connectivity of nanowire Majorana qubits relative to current superconducting-qubit architectures~\cite{bland20252d}, this limit on qubit coherence \textcolor{black}{raises doubt about the practical performance advantages of InAs-based Majorana qubits}.

{\noindent \bf Discussion.}
We have identified and characterized a previously unexplored source of decoherence in MZM-based topological qubits arising from high-frequency charge noise in the materials surrounding the device. This noise induces fluctuations in the chemical potential of the nanowire, leading to quasiparticle excitations within the bulk of the topological superconductor. These excitations produce qubit errors that are not protected by topology, increase with nanowire length, and can occur at substantial rates under currently achievable cryogenic conditions.
Our results show that although quasiparticle excitations in Majorana qubits are exponentially suppressed at sufficiently low temperatures, with current cryogenic technologies and the InAs materials platform used in~\cite{Aghaee_2023_PRB_107_23,aghaee_2025_nature_638_55}, this asymptotic regime is unlikely to be achieved in practice. While quasiparticle pair excitation arising from 1/f charge noise can be reduced by increasing the capacitance of the Majorana qubit, this simultaneously lowers the charging energy and increases susceptibility to poisoning from externally generated quasiparticles. As a consequence, the coherence of Majorana qubits is not guaranteed solely by topology, but instead 
\textcolor{black}{must be optimized by balancing} competing engineering tradeoffs.


Quasiparticle generation in alternative \textcolor{black}{nanowire} MZM architectures proposed by Microsoft such as the hexon qubit~\cite{Karzig_2017_Phys_Rev_B_95} is even more detrimental due to the six topological segments compared to the four in the tetron. Moreover, excited quasiparticles in hexons can yield both X and Z qubit errors as well as a variety of leakage errors that result in a variety of MZM configurations that are not in the qubit space (see Supplementary Sec.~\ref{supp_sec:hexon}).

In summary, we have shown that current designs on the Microsoft roadmap~\cite{Aasen2025arxiv.2502.12252} for qubits based on Majorana Zero Modes (MZMs) are expected to have substantial decoherence arising from high frequency 1/f charge noise generated by two-level fluctuators (TLFs), which can excite quasiparticle pairs in the superconducting nanowires.
Because TLF noise is ubiquitous in semiconductor materials, this mechanism persists even in defect-free nanowires operated at very low electron temperatures. Our calculations indicate that, for the current experimental implementations~\cite{aghaee_2025_nature_638_55,Aghaee.arxiv.2507.08795}, this decoherence mechanism represents a significant obstacle to achieving high-fidelity qubit operation.
Quasiparticle pair excitation arising from 1/f noise can be reduced by increasing the capacitance of the qubits, but this comes at the cost of increased susceptibility to poisoning from quasiparticles generated elsewhere in the device. As a result, optimizing the coherence of Majorana qubits requires engineering tradeoffs closely analogous to those already encountered in conventional superconducting qubit architectures. Taken together with the relatively slow gate times of the proposed architecture~\cite{Aasen2025arxiv.2502.12252}, our results suggest that this implementation of Majorana qubits may not provide a fundamental performance advantage over more mature and well-established platforms such as the transmon qubit~\cite{Koch2007p042319}.


\section*{Methods}
\label{methods}

The calculations are outlined here, with full details provided in the Supplementary Materials.

{\bf Analytic calculation of quasiparticle generation from a sudden change in chemical potential.}
Our calculations of the probability of excitation of a pair of quasiparticles from a single jump in the chemical potential from $\mu_1$ to $\mu_2$ are done by applying perturbation theory that assumes that the change in chemical potential is much smaller than the superconducting gap.
In a ring geometry the model is translationally invariant and the Hamiltonian can be written as a sum of noninteracting 2 x 2 Bogoliubov-de Gennes-Bloch Hamiltonians $H_k$ given by~\cite{Alicea_2012_IOP_75} 
\begin{equation}
H_k = \left(
\begin{array}{cc}
 \epsilon_k  &  \Delta_k  \\
 \Delta^*_k  & -\epsilon_k  \\
\end{array}
\right),
\end{equation}
where $\Delta_k$ is the superconducting gap at wavevector $k$, and $\epsilon_k$
specifies the electron energy for a given $k$ at chemical potential $\mu$
in the absence of superconductivity;
for the Kitaev chain (Eq.~\ref{eq:Kitaev}), $\epsilon_k=-w\cos(ka)-\mu.$
We calculate the probability $P_{\rm QPP}^{(1)}(k)$ of excitation of a quasiparticle pair with wavevectors $(k,-k)$, which is proportional to the square of the overlap between the ground state at $\mu_1$ with the excited state at $\mu_2$, and we then sum over the values of $k$ to obtain $P_{\rm QPP}^{(1)}$.
We then address the question of when successive transitions are essentially independent, which enables us to determine the rate of quasiparticle generation, using the numerical techniques described below.

\begin{figure}[t]
    \centering
\vskip -0.0 cm
\includegraphics[width=0.99\linewidth]{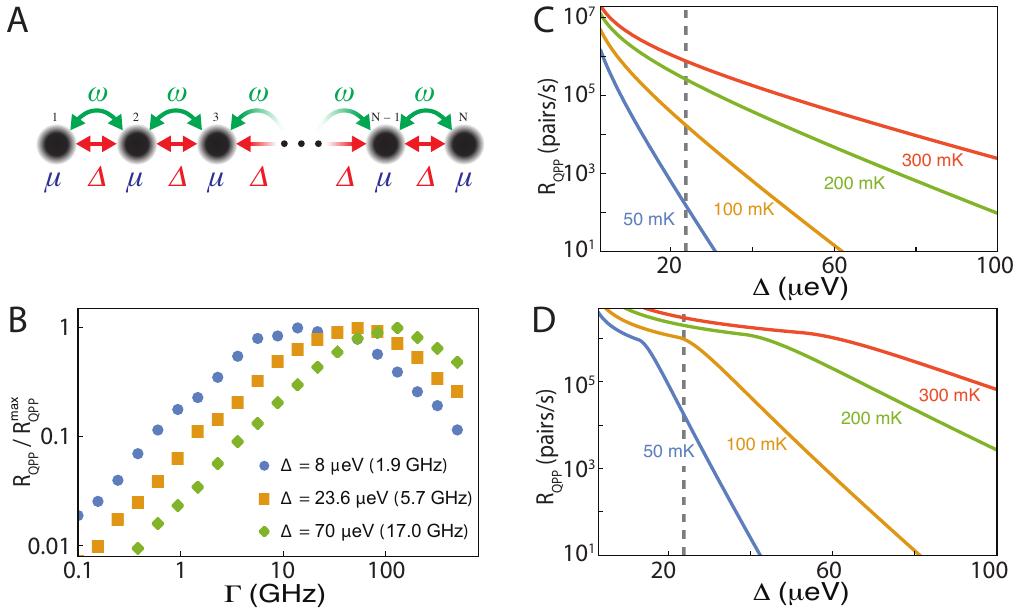}
\vskip 0cm
\caption{\small
{\bf Excitation of quasiparticle pairs by a two-level fluctuator (TLF).} 
A: Sketch of the Hamiltonian of a Kitaev chain, Eq.~\ref{eq:Kitaev}, used for numerical calculations. The electron hopping amplitude is $w$, the superconducting gap amplitude is $\Delta$, and the chemical potential is $\mu$. 
B: Numerically calculated quasiparticle pair excitation rate, $R_{\rm QPP}$, scaled by the maximum value achieved, $R_{\rm QPP}^{\rm max}$, versus TLF transition rate $\Gamma$ when the TLF effective temperature $T_{\rm eff} \gg 2\Delta$.
Here, $R_{\rm QPP}$ is maximized at $\Gamma \sim 4\Delta$. The Kitaev chain parameters used are consistent with recent experiments~\cite{aghaee_2025_nature_638_55}; see Table~1. 
C: Analytically calculated $R_{\rm QPP}$ versus $\Delta$ from Eq.~\ref{eq:R_QPP_for_finite_temperature}, evaluated at $T_{\rm eff}$ = 50~mK, 100~mK, 200~mK, and 300~mK. 50~mK is the electron temperature reported in Ref.~\cite{aghaee_2025_nature_638_55} measured on a device on which pulsed gate voltages were not applied, and 300~mK is $T_{\rm eff}$ extracted from experiments on silicon qubits (Ref.~\cite{Huang:2024p772}), which are manipulated using methods similar to those that will be used for Majorana qubits~\cite{Aasen2025arxiv.2502.12252,Karzig_2017_Phys_Rev_B_95}; see discussion in Supplementary Sec. S6. The vertical dashed line at $\Delta = 23.8~\mu$eV denotes the 90$^{\rm th}$ percentile of the gap magnitudes measured in Ref. (9) (see Supplementary Sec.~\ref{supp_sec:thermal}). For the median topological gap in Ref.~\cite{aghaee_2025_nature_638_55} of $\Delta \approx 16.8~\mu$eV, 
$R_{\rm QPP}$ exceeds 2~kHz at the reported electron temperature of 50~mK and exceeds 100~kHz at a gap of 30~$\mu$eV at a $T_{\rm eff}\sim$~190~mK. 
D: $R_{\rm QPP}$ excited by a thermally activated TLF with an attempt frequency $\Gamma_0$ of 500~GHz versus $\Delta$ at different $T_{\rm eff}$. A large $\Gamma_0$ increases $R_{\rm QPP}$ because the TLF is excited over a relatively large energy barrier, and more energy is available to excite quasiparticle pairs (see Supplementary Sec.~\ref{supplemental_subsec:attempt_frequency}).
}
\label{fig:F2}
\end{figure}

\begin{figure}[p]
\begin{center}
\vskip -1cm
\includegraphics[width=1.0\linewidth]{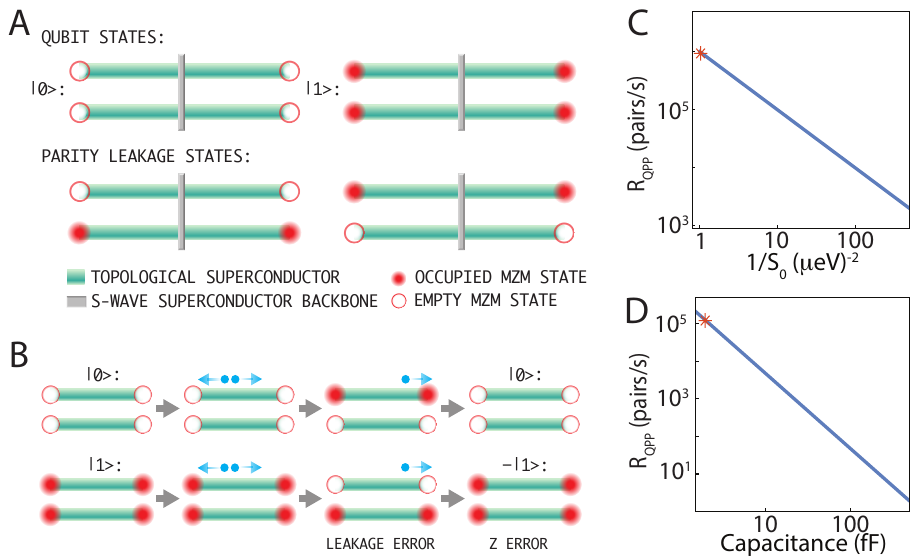}
\end{center}
\vskip -0.5cm
    \caption{\small {\bf Decoherence of tetron qubit architecture by quasiparticle pairs excited by TLF fluctuations.}  
    {A}: Schematic of a tetron qubit together with the even parity qubit states and odd parity leakage states.
    Because of the topological nature of MZMs, a single nanowire cannot have one empty and one full state.
   {B}: Decoherence of a tetron qubit by excited quasiparticle pairs.  Quasiparticles are mobile and are excited with equal and opposite momenta. MZMs absorb quasiparticles efficiently~\cite{karzig_2021_phys_rev_let_126}, so in a defect-free nanowire, the two quasiparticles in a pair are absorbed by MZMs at opposite chain ends, resulting in a qubit phase (Z) error. If the quasiparticles move diffusively, the two quasiparticles in a pair are absorbed at opposite chain ends with probability 1/3~\cite{goffage2025leakage}. There is a temporary qubit leakage error in the time interval during which one but not both of the quasiparticles has been absorbed,  but the rate of quasiparticle absorption is much higher than the rate of creation, so measurements of the parity will not exhibit a decay~(see Supplementary Sec.~\ref{supp_sec:diffusion}).
    {C-D}: Plots characterizing two strategies for improving Majorana qubit coherence, reducing the amplitude of the noise and increasing the nanowire capacitance.  
    {C}: Plot of the dependence of $R_{\rm QPP}$ on $S_0$, the coefficient specifying the amplitude of the noise power, which demonstrates that 
    reducing the density and/or amplitude of the TLF fluctuations will decrease the rate of quasiparticle pair excitation.
    {D}: Plot of the dependence of $R_{\rm QPP}$ on capacitance of the topological nanowire, demonstrating that increases of capacitance can suppress $R_{\rm QPP}$ by many orders of magnitude. 
}
  %
    \label{fig:tetron}
\end{figure}

Sec.~\ref{supp_sec:fermi_golden_rule} shows that the results for 
the rate of excitation by an ensemble of two-level fluctuators (TLFs) using the Fermi golden rule~\cite{baym2018lectures,fowler_quantum_mechanics}, which assumes that the excitation is into a continuum of quasiparticle states, 
yields 
quasiparticle excitation rates that are consistent with those yielded by the method described above that explicitly considers the time scale of decoherence of the quasiparticle pair-condensate wavefunction
when $T_{\rm eff}$, the effective temperature of the TLFs, is large enough that $k_B T_{\rm eff} > 2\Delta$, where $\Delta$ is the superconducting gap.

\ 
{\noindent \bf Numerical techniques for simulating dynamics of MZM nanowires.} 
\label{supp_sec:numerics}
Our numerical calculations have been performed using the model of a Kitaev-tetron qubit, which consists of two uncoupled Kitaev chains, where each Kitaev chain~\cite{Kitaev_2001_Physics_uspekhi_44} has the Hamiltonian given in Eq.~\eqref{eq:Kitaev}. 
This model of the Kitaev-tetron qubit has $2N$ lattice sites and therefore has a Fock space with dimension $2^{2N} \times 2^{2N}$.  As this Fock space is prohibitively large for numerical time evolution calculations even at modest chain lengths, we utilize the covariance matrix method to speed up the computation exponentially~\cite{Surace_2022_SciPost_Lec, Mishmash_2020_PRB_101}.  As in  Refs.~\cite{Mishmash_2020_PRB_101, goffage2025leakage}, we initialize the Kitaev-tetron at $t = 0$ in an equal superposition of qubit basis states
and numerically compute the probability of quasiparticle generation after each time step. A single TLF with switching rate $\Gamma$ is modeled by chemical potential fluctuating between values $\mu_{1} = 0~\mu$eV and $\mu_{2} = 0.5657~\mu$eV.
For the simulations reported in Fig.~\ref{fig:F2}, the Kitaev chain parameters used are
$w=74.88~\mu$eV, $\Delta=23.48~\mu$eV, $N  = 48$ (corresponding to a nanowire length of $\mathscr{L} = 3~\mu$m), $\mu_1=0$, and $\mu_2=0.5657~\mu$eV (see Supplementary Sec.~\ref{supp_sec:nanowire_properties} for further details).
The quasiparticle pair excitation rate shown is defined as
$R_{\rm QPP} = P_{\rm QPP,~30ns}/(30\rm{ns})$, where $P_{\rm QPP,~30ns}$ is the probability of exciting at least one quasiparticle pair after $30~\rm ns$ in one chain of the Kitaev-tetron qubit. The numerical results were averaged over 50 different trials of the TLFs.  Error bars computed from the standard deviation of the means of subsets each containing 5 different noise realizations are smaller than the size of the markers. 

{\bf Calculating the dependence of the quasiparticle pair excitation rate on the effective temperature $T_{\rm eff}$ of the two-level fluctuators (TLFs).}
The quasiparticle pair excitation mechanism considered in this paper is based on the standard model of Ref.~\cite{Dutta:1981p497} in which transitions between the two states of each TLF are stochastic, occurring at rates $\Gamma$ and 
$\Gamma e^{-\varepsilon/k_B T_{\rm eff}}$, where $\varepsilon$ is the energy asymmetry of the TLF, $k_B$ is Boltzmann's constant, and $T_{\rm eff}$ is the TLF effective temperature.
The spectral density of the noise power at nonzero angular frequency $\omega$ of a single TLF is~\cite{machlup1954noise}
\begin{equation}
    S(\omega) = \frac{1}{\pi}\frac{e^{-\varepsilon/k_B T_{\rm eff}}}
    {\left ( 1+e^{-\varepsilon/k_B T_{\rm eff}}\right )^2}
    \frac{2\Gamma}{\omega^2+4\Gamma^2}~.
    \label{eq:TLF_noise_spectrum}
\end{equation}
The frequency dependence of the noise power is Lorentzian, and the integral of the noise power becomes exponentially small when $\varepsilon/k_B T_{\rm eff} \gg 1$.
The magnitude of the noise power from an ensemble of TLFs is proportional to $T_{\rm eff}$
because the distribution of $\varepsilon$ is uniform in the standard noise model~\cite{Dutta:1981p497}.
Our formulas approximate the dependence on $\varepsilon$ by ignoring the $\varepsilon$ dependence when $\varepsilon < k_BT_{\rm eff}$ and multiplying the noise power from a TLF by the factor $e^{-\varepsilon/k_B T_{\rm eff}}$ when $\varepsilon > k_B T_{\rm eff}$.

The energy cost of exciting a quasiparticle pair is incorporated by increasing $\varepsilon$ by $2\Delta$.
Our calculations of the dependence of the quasiparticle pair excitation rate on $\Delta$ and on the TLF effective temperature also take into the effect that when $\Delta$ is changed by changing the spin-orbit length, the superconducting coherence length, which determines the extent of the Majoranas, also changes, unless the Fermi velocity in the nanowire is set to be proportional to the magnitude of the topological gap (see Supplementary Sec.~\ref{supp_sec:nanowire_properties}).
Finally, for a thermally excited TLF, the TLF must surmount an energy barrier to make a transition, which increases the amount of energy available to excite quasiparticle pairs.

To obtain an expression for the quasiparticle pair excitation rate at finite $T_{\rm eff}$, we start with Eq.~\ref{eq:quasiparticle_excitation_rate}, approximate $\mathscr{F}$ as 
$\exp(-\Gamma/(2\Delta))$, and multiply by a Boltzmann factor $\exp(-2\Delta/(k_BT_{\rm eff}$)).
We then evaluate the maximum value of $R_{\rm QPP}$ as a function of $\Gamma$:
\\
\begin{equation}
R_{\rm QPP} = \max_\Gamma 
\frac{\mathscr{L}}{16}\frac{(\delta\mu)^2\Gamma}{\hbar v_F\Delta}
\left (\exp\left [-\left (\frac{2\Delta}{k_B T_{\rm eff}}+\frac{h\Gamma}{2\Delta}\right )\right ]\right)~.
    \label{eq:R_QPP_for_finite_temperature}
\end{equation}

For a TLF that is thermally activated over an energy barrier, $\Gamma=\Gamma_0\exp(-E_b/k_B T)$, where the attempt frequency $\Gamma_0$ can be many orders of magnitude greater than $\Gamma$.
Transitions occur only when the bath provides enough energy for the TLF to surmount $E_b$, and therefore $E_b$ is available to excite quasiparticle pairs.
We account for this by changing the Boltzmann factor to $\exp \left [- (2\Delta-E_b)\Theta(2\Delta-E_b)/(k_B T_{\rm eff}) ) \right ]$, where $\Theta(x)$ is the Heaviside step function.
This procedure yields
\begin{equation}
R_{\rm QPP} = \max_\Gamma 
\frac{\mathscr{L}}{16}\frac{(\delta\mu)^2\Gamma}{\hbar v_F\Delta}
\exp\left [-\left (\max\left (0,\frac{2\Delta}{k_B T_{\rm eff}}-\frac{\Gamma_0}{\Gamma} \right )+\frac{h\Gamma}{2\Delta}\right )\right ]~,
    \label{eq:R_QPP_with_thermal_activation}
\end{equation}
where the max function is the maximum over $\Gamma$.

\newpage

%
\bibliography{alase_science} 
\bibliographystyle{sciencemag}


\section*{Acknowledgments}
We acknowledge useful conversations with Eric Bach, Jonathan Weihing, Salini Karuvade, Eric Mascot, Dimtry Pikulin, Charles Tahan, Michael Weissman, Wanli Xing, and Clare Yu.
\paragraph*{Funding:}
Work at UNSW (MCG, MCC, and SNC) was supported by the Australian Research Council, Project No.\ DP210101608 and by the Australian Research Council Centre of Excellence in Future Low-Energy Electronics Technologies (FLEET), project no.\ CE170100039, funded by the Australian government.  MCG acknowledges additional support from the Sydney Quantum Academy.
AA acknowledges support by the Australian Research Council Centre of Excellence for Engineered Quantum Systems (Grant No. CE170100009). MCC acknowledges support from a UNSW Scientia Fellowship and an Australian Research Council Discovery Early Career Research Fellowship (Grant No.\ DE240100590).
SNC acknowledges support from Google Asia Pacific Pte.\ Ltd.
\paragraph*{Author contributions:}
AA and SNC performed analytic calculations. MCG performed numerical simulations.
SNC and MCC supervised the project. All authors 
contributed equally to the preparation of the manuscript.
\paragraph*{Competing interests:}
MCC is a former employee of Microsoft and retains an equity interest in the company.
SNC owns stock in Microsoft Corporation.

\paragraph*{Data and materials availability:}
All details needed to evaluate and reproduce the results of the analytic theory in the paper are present in the paper and/or the Supplementary Materials. The computer codes used to implement all the numerical calculations reported in this paper are available in the public repository \cite{goffage_2026_zenodo_code_v3}.

\newpage

\vskip -2cm
\begin{table}[!ht]
   {\centering
    \caption{\small Table of relevant parameters. Top: Table of experimental parameters used,
    along with the source for each value.
   Middle: Table of derived parameters for the Kitaev chain Hamiltonian (Eq.~\ref{eq:Kitaev}) using the experimental parameters in the final column of the top table. 
   Bottom: Information for the Fermi velocity $v_F$ (a parameter used in Eqs.~\ref{eq:R_QPP_for_finite_temperature} and~\ref{eq:R_QPP_with_thermal_activation}), which is extracted from the parameters in the top and middle tables~(see Supplemental Sec.~\ref{supp_sec:nanowire_properties}). 
 } 
 }
    \label{tab:experimental_parameters}
   \vskip 1 cm
   \resizebox{\columnwidth}{!}{%
   \begin{tabular}{lccc}
        Quantity&  Experimental Value&  Source& Value Used Here\\
        \hline
        Nanowire length $\mathscr{L}$&  3~$\mu$m &  \cite{Aghaee_2023_PRB_107_23} and \cite{aghaee_2025_nature_638_55} & 1, 3, and 10~$\mu$m\\
        Fermi wavelength $k_{\rm F}^{-1}$&  $40-80$~nm&  ~\cite{Aghaee_2023_PRB_107_23}, Appendix A, below Eq.~1 & 40~nm  \\
        Topological Gap $\Delta$ & 10-30~$\mu$eV & \cite{aghaee_2025_nature_638_55}, Fig.~S7& from 0--200 $\mu$eV  \\
        Superconducting coherence length $\xi$ & 100 -- 250~nm & \cite{Aghaee_2023_PRB_107_23}, Sec.~2B (p 6) & 194~nm\\
        Noise power coefficient $S_0$ at reference dot &  
        0.96-1.8~$\mu$eV${}^2$&\cite{aghaee_2025_nature_638_55}, 
        Table S5  & 
        1~$\mu$eV${}^2$\\
        Charging energy $E_{C1}$ of reference dot & 180~$\mu$eV & \cite{Aghaee_2023_PRB_107_23}, Sec.~S4.2 (p 13) & 180~$\mu$eV \\
        Charging energy $E_{NW}$ of nanowire & 60~$\mu{\rm eV}$ & \cite{Aghaee_2023_PRB_107_23}, page S2 & 36~$\mu{\rm eV}$\\
        \hline
        \\
        Kitaev Chain Parameter & Derived Using &  \multicolumn{2}{c}{Value} \\
        \hline
        Kitaev chain lattice constant $a$ &$k_{\rm F}$ & \multicolumn{2}{c}{63 nm} \\
        Number of lattice sites $N$ & $k_{\rm F},~{\mathscr{L}}$ & \multicolumn{2}{c}{48, 80, and 159}\\   
        Hopping strength $w$   & $\xi,~ \Delta,~ N$ & \multicolumn{2}{c}{3.19 $\Delta$} \\ 
         Size of chemical potential jumps $\delta \mu$ & $S_0$, $E_{C1}$ & \multicolumn{2}{c}{0.5657~$\mu$eV}\\
        \hline 
        \\
        Other Derived Quantity & Derived Using & Relevant Formula & Value \\
        \hline
        Fermi velocity         & $w, ~a$       & $v_{_F} = wa/\hbar$ 
        &  ($3.05\times 10^{5})  (\Delta~{\rm in~}\mu{\rm eV})$~cm/s \\ 
        \hline 
        \\
   \end{tabular}
   }
\end{table}

	


\clearpage 

\subsection*{Supplementary Materials}
Supplementary Text\hfill\\
Figures S1 to S6\hfill\\
References \textit{(49-\arabic{enumiv})}\\ 


\newpage


\renewcommand{\thesection}{S\arabic{section}}
\renewcommand{\thesubsection}{S\arabic{section}.\arabic{subsection}}
\renewcommand{\thesubsubsection}{S\arabic{section}.\arabic{subsection}\arabic{subsubsection}}
\renewcommand{\thefigure}{S\arabic{figure}}
\renewcommand{\thetable}{S\arabic{table}}
\renewcommand{\theequation}{S\arabic{equation}}
\renewcommand{\thepage}{S\arabic{page}}
\setcounter{figure}{0}
\setcounter{table}{0}
\setcounter{equation}{0}
\setcounter{section}{0}
\setcounter{subsection}{0}
\setcounter{subsection}{0}
\setcounter{page}{1} 


\section*{Supplementary Materials for\\ \scititle}
\label{sec:supplementary_materials}

	A.~Alase,
	M.~C.~Goffage,
    M.~C.~Cassidy, and
	S.~N.~Coppersmith$^*$\\
    \and
	\small$^\ast$Corresponding author. Email: s.coppersmith@unsw.edu.au\and

\subsubsection*{This PDF file includes:}
Supplementary Text\hfil\\
Figures S1 to S6\hfil\\
\\


\newpage





\section*{Supplementary Text}

\section{Calculation of quasiparticle excitation rate 
in the TLF model of 1/f noise}
\label{supp_sec:high_T_eff}

In this section of the Supplementary Materials we 
calculate the excitation rate of quasiparticle pairs in a superconducting-semiconducting nanowire hosting Majorana zero modes (MZM) when the effective temperature of the TLFs $T_{\rm eff}$ is much larger than $2\Delta$, where $\Delta$ is the topological gap. The nanowire is subject to spatially uniform but time-dependent fluctuations in the chemical potential with a 1/f power spectral density arising from the effects of an ensemble of two-level fluctuators (TLFs). 
The effects that arise when $T_{\rm eff}$ is comparable to or smaller than $\Delta$ will be considered in Sec.~\ref{supp_sec:thermal}.

In this section, Sec.~\ref{supp_sec:one_jump} provides an analytical derivation of the quasiparticle pair excitation rate following a single jump of a TLF, using the continuum model of a 1D p-wave superconductor. Following this, in Sec.~\ref{sec:Gamma_max_calculations}, we analytically determine the excitation rate due to successive transitions of a single TLF in the regime where the quasiparticle wavefunctions completely dephase between successive TLF transitions. We provide bounds on this rate depending on two limiting mechanisms for the wavefunction dephasing. 
Sec.~\ref{supp_sec:delta_dependence} presents numerical results showing that $\Gamma_{\rm max}$ increases with the superconducting gap. 
In Sec.~\ref{supp_sec:numerics_details} we calculate numerically the quasiparticle pair excitation rate in the regime where the quasiparticle wavefunctions do not completely dephase between TLF transitions.
Finally, in Sec.~\ref{supp_sec:delta_mu_derivation} we provide a derivation of the extraction of the TLF fluctuation amplitude as used in our models from experimentally measured 1/f charge noise power spectra.

\subsection{Quasiparticle excitation rate from one sudden 
change of the chemical potential} \label{supp_sec:one_jump}

The numerical calculations reported in this paper are performed for the Kitaev chain, but the mechanism that gives rise to quasiparticle pair excitation in the bulk of a superconductor applies to a broader class of models.
We choose to
calculate the quasiparticle pair excitation rate 
using a continuum model of a 1D $p$-wave superconductor (see, e.g., Eq.\ 10 in~Ref.~\cite{Leijnse_2012}):
\begin{equation}
    \hat{H} = \int dx \left[\Psi^\dagger(x)
    \left(\frac{p_x^2}{2m}-\mu\right)\Psi(x) +
    \Psi(x)\Delta_0 p_x\Psi(x) + \text{H.c.}\right],
\end{equation}
where $\Psi^\dagger(x)$ creates an electron at position $x$,
$p_x$ is the momentum operator in $x$ direction (along the wire),
$\mu$ is the chemical potential, $m$ is the effective mass
of an electron, and $\Delta_0$ is the superconducting 
pairing parameter that is 
assumed to be positive without loss of generality.
We calculate the excitation rate with periodic boundary conditions applied, which yields the bulk contribution.
This system is translationally invariant, so momentum is a good quantum number and $\hat{H}$
can be decomposed into sectors labeled by wavevector $k$, with
the corresponding 2 x 2 Bogoliubov-de-Gennes-Bloch Hamiltonian $H_k$ in the basis 
$(c_k \; c_{-k}^\dagger)^{\rm T}$ given by~\cite{Alicea_2012_IOP_75}
\begin{equation}
H_k = \left(
\begin{array}{cc}
 \epsilon_k  &  \Delta_k  \\
 \Delta^*_k  & -\epsilon_k  \\
\end{array}
\right).
\end{equation}
Here, $\Delta_k$ is the superconducting gap at wavevector $k$, and $\epsilon_k = \hbar^2k^2/2m - \mu$ 
specifies the electron energy as a function of $k$ in the absence of superconductivity. 
Due to parity conservation, the quasiparticles are excited in pairs, so the 
unique ground state $\ket{\Omega}$ is excited predominantly to states of the form
$\ket{\eta_{k}} = d_{k}^\dagger d_{-k}^\dagger \ket{\Omega}$, where \textcolor{black}{$d_{k}^\dagger$ denotes the creation operator} of a quasiparticle with wavevector $k$
for the initial value of the chemical potential.
To obtain the probability $P_{QPP}^{(1)}$ that one jump of $\mu$ excites a quasiparticle pair, we calculate the probability $P_{\rm QPP}^{(1)}(k)$ of excitation of a quasiparticle pair with wavevectors $(k,-k)$ and then sum over all such pairs. 

Let $\ket{\Omega'}$ denote the ground state after the sudden change in chemical potential.
Using perturbation theory \textcolor{black}{for small $\delta\mu$}, we express $\ket{\Omega'}$ as
\begin{equation}
\label{eq:perturbation}
    \ket{\Omega'} \approx \ket{\Omega} + \sum_{k \in {\rm BZ}} \beta_k\ket{\eta_k},
    \quad \beta_k = \delta\mu\frac{\bra{\eta_k}\hat{N}\ket{\Omega}}{2E_k},
\end{equation}
where $E_k = \sqrt{\epsilon_k^2+|\Delta_k|^2}$ and the sum is over $k$ in the first Brillouin zone.
The quasiparticle annihilation operators can be expressed in terms of the operators $c_k^\dagger$ and $c_k$ that create and annihilate a fermion with wavevector $k$ as~\cite{Alicea_2012_IOP_75} 
\begin{equation}
\label{eq:ukvk}
d_k = u_k c_k + v_k c_{-k}^{\dagger}, \quad
{\rm~with~}
u_k = \frac{\Delta_k}{\abs{\Delta_k}}\sqrt{\frac{E_k + \epsilon_k}{2E_k}}, \quad
v_k = \frac{E_k-\epsilon_k}{\Delta_k}u_k.
\end{equation}
Expressing the number operator as
$\hat{N} = \sum_{k \in {\rm BZ}}c_k^\dagger c_k$ yields 
\begin{equation}
\expval{\eta_k|\hat{N}|\Omega} 
= \expval{\eta_k|\sum_{k' \in {\rm BZ}}c_{k'}^\dagger c_{k'}|\Omega}
= \sum_{k' \in {\rm BZ}}\expval{\Omega|d_{-k}d_k c_{k'}^\dagger c_{k'}|\Omega}.
\end{equation}
For $k'\ne \pm k$, the summand is $\expval{\eta_k|\hat{N}_{k'}|\Omega} 
    = \expval{\Omega|c_{k'}^\dagger c_{k'}d_{-k}d_k|\Omega} = 0$.
Therefore, we get
\begin{equation}
\label{eq:overlap2}
    \expval{\eta_k|\hat{N}|\Omega} = \expval{\eta_k|c_{k}^\dagger c_{k}|\Omega} + \expval{\eta_k|c_{-k}^\dagger c_{-k}|\Omega}.
\end{equation}
Using Eq.~\eqref{eq:ukvk}, we obtain
\begin{equation}
    \expval{\eta_k|c_k^\dagger c_k|\Omega}
    = \expval{\Omega|d_{-k}d_k (u_k d_k^\dagger+v_{-k}^* d_{-k} )
    (u_k^* d_k+v_{-k} d_{-k}^\dagger )|\Omega} = u_kv_{-k}.
\end{equation}
Similarly, we get $\expval{\eta_k|c_{-k}^\dagger c_{-k}|\Omega}
    = -u_{-k}v_{k}$, and substituting in Eq.~\eqref{eq:overlap} yields
\begin{equation}
\label{eq:overlap}
\expval{\eta_k|\hat{N}|\Omega} = u_kv_{-k}  - u_{-k}v_k
= \frac{\Delta_k}{E_k}. 
\end{equation}
Here we used $\epsilon_{-k} = \epsilon_{k}$ and $\Delta_{-k} = -\Delta_{k}$
and expressions for $u_k$ and $v_k$ in Eq.~\eqref{eq:ukvk}.
By substitution in Eq.~\eqref{eq:perturbation}, we get
\begin{equation}
    \beta_k = \frac{\delta\mu\expval{\eta_k|\hat{N}|\Omega}}{2E_k} 
    = \frac{\delta\mu\Delta_k}{2E_k^2}.
\end{equation}
Finally, the probability of excitation of a $(k,-k)$ quasiparticle pair is obtained to be
\begin{equation}
\label{eq:excitation_for_one_k}
    P_{\rm QPP}^{(1)}(k)=\abs{\beta_k}^2 = \frac{\delta\mu^2\left |\Delta_k^2\right | }{4E_k^4}.
\end{equation}
In Eq.~\eqref{eq:excitation_for_one_k} the probability of exciting a quasiparticle pair is proportional to the square of the chemical potential change, and the factor of $E_k^4$ in the denominator
ensures that only $k$-vectors whose energies are of order $\Delta_k$ 
contribute significantly to the sum over $k$-vectors.

Integrating the contributions over wavevectors $k$ 
yields the total probability of excitation of a pair of quasiparticles by a single jump of the chemical potential, $P_{\rm QPP}^{(1)}$. 
In the experimentally relevant regime, the integrand 
is strongly peaked near the Fermi wavevector $k_{\rm F}$ defined by 
$\epsilon_{k_{\rm F}}=0$. Therefore, we can use a linearized 
energy dispersion~\cite{spaanslatt2015topological}, 
$\epsilon_k = \hbar v_{\rm F} |k|-\mu$, 
where $v_{\rm F} = (1/\hbar)\partial \epsilon_k/\partial k$ is the Fermi velocity of the electrons in the nanowire and $\hbar=h/2\pi$ is the reduced Planck constant.
We can also
approximate $\Delta_k \approx \Delta_{k_{\rm F}}$.
In the limit of small excitation rate, we get
\begin{align}
\label{eq:excitation_for_one_TLF}
    P_{\rm QPP}^{(1)} &= 
    \frac{\mathscr{L}}{2\pi}\int_{0}^\infty
    dk~\frac{1}{4}\left | \frac{\delta\mu}{\Delta_{k_{\rm F}}} \right |^2\frac{\Delta_{k_{\rm F}}^4}{(\Delta_{k_{\rm F}}^2+(\hbar v_{\rm F}k-\mu)^2)^2}
    \nonumber\\
    &\approx
    \frac{\mathscr{L}}{2\pi}\int_{-\infty}^\infty
    dk~\frac{1}{4}\left | \frac{\delta\mu}{\Delta_{k_{\rm F}}} \right |^2\frac{\Delta_{k_{\rm F}}^4}{(\Delta_{k_{\rm F}}^2+(\hbar v_{\rm F}k-\mu)^2)^2}
    \nonumber\\
    &= \left | \frac{\delta\mu}{\Delta_{k_{\rm F}}} \right |^2
    \frac{\mathscr{L}}{16}\frac{\Delta_{k_{\rm F}}}{\hbar v_{\rm F}},
\end{align}
where $\mathscr{L}$ is the sample length.

\subsection{Quasiparticle excitation rate from a single two-level fluctuator with transition rate $\Gamma$}
\label{sec:Gamma_max_calculations}
The rate at which quasiparticle pairs are excited by the effects of TLFs depends on whether each transition of a TLF causes more quasiparticles to be excited.
If the TLF transition rate $\Gamma$ is low, then the effects of successive transitions are uncorrelated, and each TLF transition induces the production of more quasiparticles.
This argument yields the rate of
quasiparticle pair excitation due to a single TLF to be
$R_{\rm QPP} = \Gamma P_{\rm QPP}^{(1)}$. 
However, if the chemical potential returns back to its original value before the wavefunction has had a chance to evolve at all, then the return transition actually would reduce the number of excited quasiparticles.
We find numerically that for switching rates
comparable to or higher than $\left |\Delta_{k_{\rm F}}\right |/h$, the rate of
quasiparticle pair excitation is
\begin{equation}
    R_{\rm QPP} = \mathscr{F}\Gamma P_{\rm QPP}^{(1)}
    = \mathscr{F}\Gamma
   \frac{\delta\mu^2}{\left | \Delta_{k_{\rm F}}\right |}
    \frac{\mathscr{L}}{16\hbar v_{\rm F}}~,
    \label{eqn:qpp_rate_mult_fact}
\end{equation}
with $\mathscr{F} \in [0,1]$ a multiplicative factor.

In the regime that successive TLF transitions excite more quasiparticle pairs, the rate of excitation is proportional to $\Gamma$.
The excitation rate again becomes small when $\Gamma$ is much larger than $2\Delta$, so the excitation rate of quasiparticle pairs reaches a maximum value at a value of $\Gamma$ that we will denote as $\Gamma_{\rm max}$.
In this section we present methods for obtaining bounds for $\Gamma_{\rm max}$.
In the main text we present the results of numerical investigations of Kitaev chains with a time-varying chemical potential but no other source of dissipation, such as coupling to phonons or impurities.
Because the Kitaev model lacks some of the dissipative mechanisms that are present in real nanowires, we expect our numerical results to overestimate the degree of quantum coherence and thus to yield an underestimate of the number of quasiparticle pairs excited.
Nonetheless, as shown in Fig.~2 in the main text,
the number of quasiparticles excited by a TLF within the Kitaev model continues to increase up to frequencies of order of the superconducting gap $\Delta$.

Here we derive analytic bounds for $\Gamma_{\rm max}$.
Our analytic lower bound to $\Gamma_{\rm max}$ is obtained by following Refs.~\cite{hu2011two,gamble2012two} and noting that the coherent superposition is lost when a quasiparticle collides with an impurity or with the sample boundary.
We also note that,
as pointed out in Ref.~\cite{karzig_2021_phys_rev_let_126}, when a quasiparticle reaches the end of a topological nanowire, it interacts strongly with the Majorana Zero Mode (MZM) at the end of the wire, and the MZM mediates the decay of the quasiparticle.
This quasiparticle decay process is not included in the Hamiltonian of the Kitaev chain, which has no dissipative processes, but the process is present in real nanowires via coupling between the quasiparticles and phonons.
In the physical system, a state that is a superposition of the ground state and an excited quasiparticle pair state decoheres when a quasiparticle reaches the end of the wire.
The time until a scattering event is longest when the only scattering is at the sample boundary, when the quasiparticles travel ballistically to the boundary at the Fermi velocity $v_{\rm F}$.
At least one quasiparticle reaches the boundary within a time $\mathscr{L}/(2 v_{\rm F})$, where $\mathscr{L}$ is the sample length, so this argument yields $2 v_{\rm F}/\mathscr{L}$ as an estimate for the 
lower bound on $\Gamma_{\rm max}$.
Because the quasiparticle decays when it interacts with the MZM at the chain end, we expect the factor $\mathscr{F}$ to be close to unity, so we find the bound
\begin{equation}
    R_{\rm QPP} \gtrsim \frac{(\delta\mu)^2}{4\hbar\left |\Delta_{k_{\rm F}}\right |}~.
\end{equation}


An upper bound to $R_{\rm QPP}$ is obtained by noting that the energy difference between the condensate pair and the quasiparticle pair is $2 E_k$, where $E_k^2 = \left | \Delta_k\right |^2 + (\hbar v_{\rm F}k-\mu)^2$.
Therefore,
the relative phase of the two terms in the time-dependent wavefunction
\begin{equation}
    |\psi_k(t)\rangle = A|\Omega\rangle + Be^{2iE_kt/\hbar}|\eta_k\rangle
\end{equation}
is near zero so long as $2 E_k/(\hbar \Gamma) \ll 2\pi$.
Because the quasiparticle formation itself is dominated by $k$'s for which $E_k$ is of order $\Delta_{k_{\rm F}}$, this argument yields an upper bound to $\Gamma_{\rm max}$ of order $2\Delta_{k_{\rm F}}/\hbar$.
As discussed in the next two subsections, our numerical investigations of Kitaev chains are consistent with a value of $2\Gamma_{\rm max}$ close to this upper bound of $\Delta/h$, but with a rate of quasiparticle pair formation that is somewhat below that obtained if successive TLF transitions were entirely independent.
We find that the maximum rate of quasiparticle pair formation can be written
\begin{equation}
    R_{\rm QPP} \lesssim \frac{\mathscr{LF}}{4}\frac{(\delta\mu)^2}{\hbar^2 v_{\rm F}},
\end{equation}
\textcolor{black}{where the numerical factor $\mathscr{F}$ is between $1/2$ and $1$ when $\Gamma \approx 4 \left |\Delta\right |/h$ (it is somewhat gap-dependent).}


Thus we have obtained bounds on the rate of quasiparticle pair formation:
\begin{equation}
    \frac{(\delta\mu)^2}{4\hbar\left |\Delta_{k_{\rm F}}\right |} \lesssim R_{\rm QPP} \lesssim  \frac{\mathscr{LF}}{4}\frac{(\delta\mu)^2}{\hbar^2 v_{\rm F}}.
    \label{eq:quasiparticle_rate}
\end{equation}

It turns out that for the values of the relevant quantities appropriate for the experiments reported in Ref.~\cite{aghaee_2025_nature_638_55}, the upper and lower bounds for $R_{\rm QPP}$ are of the same order of magnitude.

We have further characterized the dependence of the quasiparticle excitation rate on the TLF frequency by performing numerical simulations of Kitaev chains with open boundary conditions that are presented in detail in Sec.~\ref{supp_sec:numerics_details}.  
Our numerical results for the Kitaev chain are consistent with the upper limit in Eq.~\ref{eq:quasiparticle_rate}, yielding a rate of quasiparticle pair excitation that is proportional to the nanowire length. 

Finally, we note that the total rate of quasiparticle excitation is dominated by the effects of the fastest TLF for which the effects of successive transitions add.  This is because the rate of quasiparticle excitation by a TLF of frequency $\Gamma$ is proportional to $\Gamma$, and in a 1/f spectrum the values of $\Gamma$ are logarithmically distributed in frequency~\cite{Dutta:1981p497}.


\subsection{Numerical calculations of the quasiparticle pair excitation rate on the magnitude of the superconducting gap}
\label{supp_sec:delta_dependence}
In this subsection we present numerical results characterizing the dependence of the quasiparticle pair excitation rate on nanowire parameters in the limit of large $T_{\rm eff}$.
Our numerics demonstrate that the quasiparticle pair excitation rate grows linearly with system size, with a size-independent correction arising from the finite extent of the Majorana modes at the chain ends.
As we have shown in the main text Fig.~\ref{fig:F2}, the TLF switching rate $\Gamma$ at which the maximum number of quasiparticle pairs are generated, $\Gamma_{\rm max}$, grows linearly with the magnitude of the superconducting gap $\Delta$. 
Fig.~\ref{fig:supp_gamma_max_factor} shows that $\Gamma_{\rm max}$ is independent of $\mathscr{L}$, so that it is sufficient to consider a single nanowire length, which we take to be $\mathscr{L} = 3$~$\mu$m~\cite{aghaee_2025_nature_638_55,Aghaee.arxiv.2507.08795}. As shown in Fig.~\ref{fig:F2} in the main text,
the TLF transition rate at which the quasiparticle pair excitation rate $R_{\rm QPP}$ grows as $\Delta$ is increased and reaches a maximum at 
a value proportional to $\Delta$.


 \subsection{Numerical calculations using the Kitaev chain Hamiltonian of the quasiparticle pair excitation rate induced by a single TLF at large $T_{\rm eff}$} 
 \label{supp_sec:maximising_R_QPP_versus_Gamma}
\begin{figure}[!ht]
    \centering
    \includegraphics[width=1\linewidth]{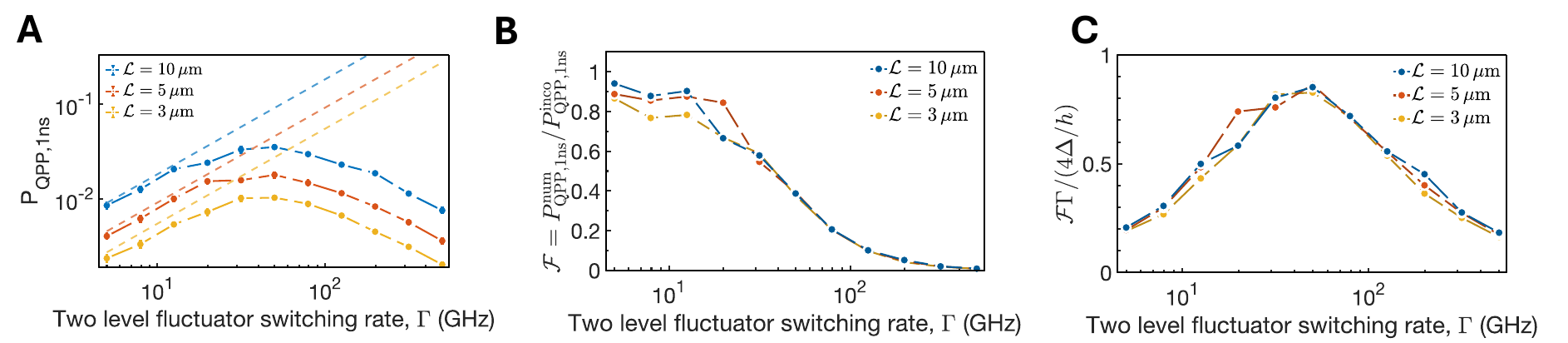}
    \caption{\small {\bf Dependence of rate of excitation of quasiparticle pairs (QPPs) on nanowire length.}  These calculations were done with a single two-level fluctuator (TLF) in Kitaev chains with lengths $\mathscr{L} = 3\;\unit{\um},~5\;\unit{\um},~10\;\unit{\um}$, hopping parameter $w = 74.88$\;\unit{\micro\eV}, superconducting gap amplitude $\Delta = 23.48$\;\unit{\micro\eV}, and one TLF switching between the chemical potential values $\mu_1 = 0$\;\unit{\micro\eV} and $\mu_2 = 0.5657$\;\unit{\micro\eV} at different transition rates $\Gamma$. 
    \textbf{A}: Numerically calculated probability of exciting at least one QPP in a Kitaev chain over $1~\rm{ns}$, $P_{\rm QPP,~1~ns}^{\rm{num}}$ (solid markers with dashed-dot lines).
    The dashed lines show $P_{\rm QPP}^{\rm inco}$, the probabilities that would be obtained if the dynamics were completely incoherent and the QPP generation of successive transitions of the TLF were completely independent. \textbf{B:} Plot of the ratio $\mathscr{F} = P_{\rm QPP}^{\rm num}/P_{\rm QPP}^{\rm inco}$ (see Eq.~\ref{eqn:qpp_rate_mult_fact}) versus TLF transition rate $\Gamma$.
    Note that the ratio $P_{\rm QPP}^{\rm num}/P_{\rm QPP}^{\rm inco} \rightarrow 1$ as $\Gamma \to 0$.
    \textbf{C:}  Plot of $\mathscr{F}\Gamma$ scaled by $4 \Delta/h$.
    The dependence of $\mathscr{F}$ on nanowire length $\mathscr{L}$ is extremely weak. In the regime where $R_{\rm{QPP}}$ is near its maximum, for this value of $\Delta$ the multiplicative factor is $\mathscr{F}\Gamma \approx 0.85 \times 4\Delta/h$. This value is used in Eq.~\eqref{eq:quasiparticle_excitation_rate_intermediate} to obtain the estimate of the rate of excitation of quasiparticle pairs in the limit of large $T_{\rm eff}$. 
    } 
\label{fig:supp_gamma_max_factor}
\end{figure}

In this subsection we present numerical calculations used to determine the maximum rate of quasiparticle pair excitation as a function of the TLF transition rate $\Gamma$, or equivalently, the behavior of the multiplicative factor $\mathscr{F}$ present in Eq.~\eqref{eqn:qpp_rate_mult_fact} in the limit of large $T_{\rm eff}$. 
Again, the factor $\mathscr{F}$ accounts for the fact that the effects of successive jumps of a TLF are not completely uncorrelated between jumps when the TLF transition rate is high.
We would expect that for sufficiently low $\Gamma \ll \Delta/h$ the coherent superposition of superconducting condensate and excited quasiparticle pairs completely dephases between consecutive TLF transitions and $\mathscr{F} = 1$, while for sufficiently high $\Gamma \gg \Delta/h$ the superposition of condensate and quasiparticle pairs experiences negligible change between consecutive TLF transitions and $\mathscr{F} \rightarrow 0$. 
Fig.~\ref{fig:supp_gamma_max_factor} presents numerical calculations that support this claim and estimate $\mathscr{F}$ at intermediate $\Gamma \sim \Delta/h$. Fig.~\ref{fig:supp_gamma_max_factor}a presents the numerically computed $R_{\rm QPP}$ along with $\Gamma P_{\rm QPP}^{(1)}$, which are in agreement for small $\Gamma$. 
Fig.~\ref{fig:supp_gamma_max_factor}b presents the ratio $\mathscr{F} = R_{\rm QPP}/\Gamma P_{\rm QPP}^{(1)}$ which clearly transitions from $1$ at small $\Gamma$ to $0$ at large $\Gamma$. Furthermore, Fig.~\ref{fig:supp_gamma_max_factor}c presents $\mathscr{F}\Gamma/(4\Delta/h)$ as a function of $\Gamma$, demonstrating that for intermediate $\Gamma$ where $R_{\rm{QPP}}$ is at its maximum, $\mathscr{F}\Gamma \approx 0.7 \times 4\Delta/h$. 
Panels B and C of Fig.~\ref{fig:supp_gamma_max_factor} show that the numerical factor $\mathscr{F}$ does not have a significant dependence on $\mathscr{L}$, so the quasiparticle pair excitation rate ($R_{\rm QPP}$) is proportional to the nanowire length.
As the TLF transition 
rate $\Gamma$ increases, $R_{\rm QPP}$ increases proportionally to $\Gamma \lesssim 4\Delta/h$.
\textcolor{black}{For $\Gamma \gtrsim 4\Delta/h \sim 100$\;\unit{\GHz}, $R_{\rm QPP}$ 
plateaus at a maximum, and 
then decreases as $\Gamma$ increases further (Fig.~\ref{fig:supp_gamma_max_factor}c)}.
The value of $\mathscr{F}\Gamma$ at the plateau
is found to be in the range 
$(\mathscr{F}\Gamma)_{\rm max} \approx (0.7-0.85) \times 4\Delta/h$, with a slight dependence on the magnitude of $\Delta$.
For definiteness, we take $(\mathscr{F}\Gamma)_{\rm max}$, yielding
a maximum rate of quasiparticle pair excitation by one TLF, 
$R_{\rm QPP} = R_{\rm QPP, max}$, of
\begin{equation}
    R_{\rm QPP, max} \approx \mathscr{L}\frac{0.7}{8}\frac{(\delta\mu)^2}{ \pi \hbar^2v_{\rm F}}~.
    \label{eq:quasiparticle_excitation_rate_intermediate}
\end{equation}

In the succeeding subsection~\ref{supp_sec:delta_mu_derivation} we present our method to extract the TLF fluctuation amplitude as used in our models from experimental measurements of the 1/f charge noise power spectra.

\subsection{Extraction of TLF fluctuator amplitude $\delta\mu$ from the experimentally measured power spectrum}
\label{supp_sec:delta_mu_derivation}

 In this subsection we will focus on the chemical potential and the TLF chemical fluctuation magnitude $\delta\mu$ to the parameter $S_0$, where the chemical potential noise power spectrum is $S(\omega) = S_0/\omega$.
Two slightly different methods to extract $\delta\mu$ from $S_0$, the coefficient of the measured spectral density of fluctuations of the chemical potential from Ref.~\cite{aghaee_2025_nature_638_55}, listed in Table~1 in the main text as $S_0/\omega$, where $\omega$ is the angular velocity, $\omega= 2\pi f$, with $f$ the frequency in Hz. 
A relation between experimentally measured 
$S_0$ and $\delta \mu$ can be obtained 
by first constructing an ensemble of TLFs 
that gives rise to 1/f noise spectrum~\cite{Dutta:1981p497}.
Consider a TLF that contributes to the change in chemical potential
by amplitude $\delta\mu$ and has switching rate $\Gamma$.
The density of spectral power due to an individual symmetric TLF is Lorentzian~\cite{machlup1954noise},
\begin{equation}
    S_\Gamma(\omega) 
    =\frac{(\delta\mu)^2}{2\pi}\frac{\Gamma}{4\Gamma^2+\omega^2}, 
    \quad \omega \ne 0.
\end{equation}
Let $D(\Gamma)$ denote the density of TLFs at $\Gamma$.
A 1/f spectrum results if the values of $\Gamma$ are equally spread on a logarithmic scale, in other words, if the number of values of $\Gamma$ is the same in each 
decade of frequency~\cite{Dutta:1981p497}.
To replicate a $1/\omega$ dependence, we 
choose $D(\Gamma) = 1/\Gamma$.
This distribution ensures that there is one TLF  
with lifetime $\Gamma \in [\Gamma_0 , e\Gamma_0]$
for any $\Gamma_0>0$, since
\begin{equation}
    \int_{\Gamma_0}^{e\Gamma_0}D(\Gamma)d\Gamma = 1.
\end{equation}
Therefore, this distribution is equivalent to having 
a set of TLFs with switching rates given by $\{\Gamma_0 e^n,\ n \in \mathbb{Z}_+\}$,
where $\Gamma_0>0$ is a low-frequency cutoff.
Using this density of TLFs, we obtain the total power spectral density
to be
\begin{equation}
    S(\omega) = \int_{0}^{\infty} D(\Gamma)S_\Gamma(\omega) d\Gamma = 
    \frac{\delta\mu^2}{8\omega} =: \frac{S_0}{\omega},
    \label{eq:S_0_mu_relation}
\end{equation}
so we obtain $S_0 = \delta\mu^2/8$.

Now we account for the fact that Ref.~\cite{aghaee_2025_nature_638_55} reports the magnitude of the chemical potential fluctuations at a reference dot that has a capacitance that differs from that of the nanowire.
This must be taken into account because
the magnitude of the jumps of the chemical potential of the nanowire also depends on the ratio of the dot capacitance to the nanowire capacitance~\cite{Yu.arxiv.2512.05644}.
This dependence on capacitance arises because the noise is the result of charge motion, and changing the charge on a capacitor with capacitance $C$ by an amount $\delta q$ changes the voltage on the capacitor by $\delta q/C$.
Therefore, the size of the chemical potential jump is proportional to $1/C$.

The value of the 
coefficient $S_0^{\rm dot}$ for the reference dot in Ref.~\cite{aghaee_2025_nature_638_55} 
is measured to be $\approx (1\;\unit{\micro\eV})^2$,
which, from Eq.~\ref{eq:S_0_mu_relation}, corresponds to 
$\delta\mu^{\rm dot} = 2\sqrt{2S_0^{\rm dot}} = 
2.83$\;\unit{\micro\eV}. 
To relate the value of $\delta\mu$ at the reference 
dot to its value at the nanowire, we assume that 
the magnitudes of the charge motions are the same, 
so that
\begin{equation}
    \delta\mu^{\rm nanowire} = \delta\mu^{\rm dot}/(C^{\rm nanowire}/C^{\rm dot}).
\end{equation}
Writing this in terms of the charging energy $E_C = e^2/2C$, we find
\begin{equation}
    \delta\mu^{\rm nanowire} = \delta\mu^{\rm dot}(E_C^{\rm nanowire}/E_C^{\rm dot}).
\end{equation}
The charging energy of the reference dot in Ref.~\cite{aghaee_2025_nature_638_55} is reported to be 160 $\mu$eV.
Based on the discussion in Ref.~\cite{aghaee_2025_nature_638_55}, we estimate the smallest feasible charging energy consistent with the Microsoft roadmap~\cite{Aasen2025arxiv.2502.12252} to be about 5 times smaller, or about 30~$\mu$eV (which corresponds to a temperature of 0.35~K).
Thus we find
\begin{equation}
    \delta\mu^{\rm nanowire} = \delta\mu^{\rm dot}(E_C^{\rm nanowire}/E_C^{\rm dot}) = 2.83/5 = 0.566~\mu{\rm eV}~.
\end{equation}

\begin{figure}[!ht]
    \centering
    \vskip -2cm
\includegraphics[width=1.0\linewidth]{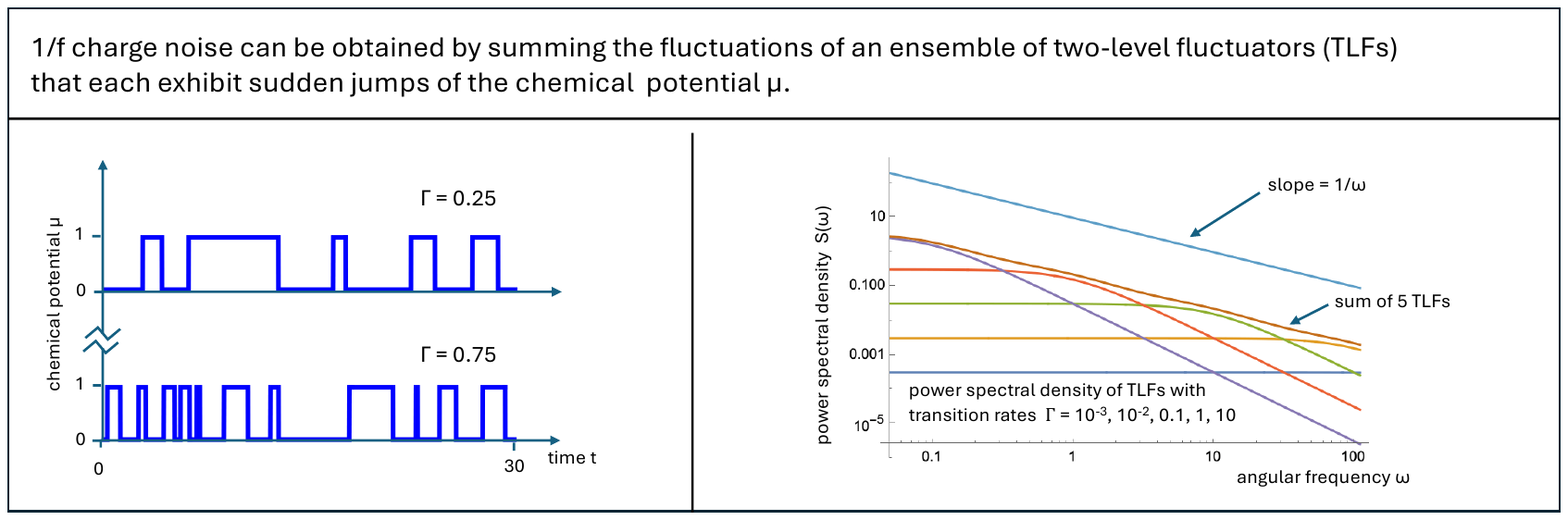}
    \vskip -5.0cm
    \caption{\small {\bf Construction of 1/f noise as the sum of Lorentzians from an ensemble of two-level fluctuators (TLFs).}
    Left: Plot of changes in chemical potential due to switches of TLFs versus time for TLFs with different transition rates $\Gamma$.
    Right: Illustration of how 1/f noise emerges from the fluctuations of an ensemble of TLFs in which the frequencies are uniformly distributed in the logarithm of the frequency, in other words, a constant density of $\Gamma$'s per decade of frequency.
    In this illustration, there is one $\Gamma$ per decade of the angular frequency $\omega = 2\pi f$.  For this case, the total spectral density obtained by summing over all the TLFs in between 1.44 and 2.24 times the spectral density of a single TLF.
    \\
    \\
    \label{fig:adding_fluctuators_supp}}
\end{figure}

We now present another method to extract a lower bound on the
magnitude of $\delta \mu$ from $S_0$.
We analyze the noise power spectral density of a specific discrete ensemble of TLFs to relate $S_0$ to $\delta \mu$. Fig.~\ref{fig:adding_fluctuators_supp} illustrates that adding the fluctuation power spectral density from a rather small number of TLFs (one per decade in angular frequency $\omega$) yield a total spectral power density that is quite close to 1/f.
For this case, the spectral power density at an angular frequency $\omega$ is due mostly from a TLF with a value of $\Gamma$ that is near $\omega$---the fraction of the total spectral power density from the TLF with the closest $\Gamma$ varies from 0.69 to 0.446.
Therefore, we can obtain a lower bound on a possible value of $\delta \mu$ by setting the spectral power density at frequency $\omega$ of a single TLF to 0.446 $S_0/\omega$.
Since the spectral power density $S(\omega)$ of a single TLF is $(\Gamma\delta\mu^2/2\pi)/(4\Gamma^2+\omega^2)$~\cite{machlup1954noise}, we can choose to look at a single TLF with transition frequency $2\Gamma=\omega$ and find that the spectral power density from that TLF is $(\delta\mu)^2/(4\pi\omega)$.
Since the spectral power from that one TLF would be a fraction of the total power that is between 1.44 and 2.24, we have
\begin{equation}
    \frac{8\pi S_0}{2.24}<(\delta\mu)^2 <
    \frac{8\pi S_0}{1.44}~.
\end{equation}
This result is consistent with the result obtained in Eq.~\ref{eq:S_0_mu_relation} by integration over a continuous density of TLF transition rates.

\section{Calculation of quasiparticle excitation rate using
Fermi's golden rule}
\label{supp_sec:fermi_golden_rule}
In this section of the Supplementary Materials we provide alternate analytical derivations for the quasiparticle pair excitation rate induced by a single TLF and by an ensemble of TLFs (Sec.~\ref{supp_sec:fermi_ensemble_TLF}), using Fermi's golden rule. 
The results from Fermi's golden rule agree well with those from summing the contributions from an ensemble of TLFs in the limit of large TLF effective temperature $T_{\rm eff}$.
We note that calculations using Fermi's golden rule of quasiparticle excitation rates in the presence of noise have been performed previously for spatially uniform noise for excitations from MZM modes in the presence of noise with a Lorentzian power spectrum~\cite{PhysRevB.88.075431} and with a Gaussian power spectrum~\cite{Mishmash_2020_PRB_101}. 
The derivation of Fermi's golden rule involves a somewhat opaque limiting procedure~\cite{baym2018lectures,fowler_quantum_mechanics}, but the results agree well with the analysis above that sums the effects of individual TLFs along with numerical studies of the Kitaev chain.
The golden rule formalism yields an analytical result for the multiplicative factor $\mathscr{F}$ in Eq.~\ref{eq:R_QPP_with_F_included} that agrees well with the numerical results in Sec.~\ref{supp_sec:maximising_R_QPP_versus_Gamma}. The golden rule method also gives a pair excitation rate for an ensemble of TLFs consistent with 1/f charge noise in the limit of large $T_{\rm eff}$ that is consistent with the results from the calculations reported in the main text.

\subsection{Quasiparticle excitation rate due to a single TLF 
using Fermi's golden rule} \label{supp_sec:fermi_single_TLF}
In this section, we calculate the $\mathscr{F}$ factor 
analytically using Fermi's golden rule.
Fermi's golden rule for fluctuating noise~\cite{weinberg2015lectures}
states that the rate of generation of quasiparticle pairs at wavevector $k$ is 
\begin{equation}
\label{eq:fermigoldenrule}
R_{\rm QPP}(k) = \frac{2\pi}{\hbar^2}\abs{\expval{\eta_k|\hat{N}|\Omega}}^2 S(2E_k/\hbar)
\end{equation}
where $E_k = \sqrt{\epsilon_k^2 + \abs{\Delta_k}^2}$ is the bulk quasiparticle energy
and $\ket{\eta_k} = d^\dagger_k d^\dagger_{-k}\ket{\Omega}$ is the state 
with excited $(k,-k)$ pair of quasiparticles. 
For a single TLF with switching rate $\Gamma$, the noise spectral density is~\cite{machlup1954noise}
\begin{equation}
S(\omega) = \frac{\delta\mu^2}{2\pi}\frac{\Gamma}{4\Gamma^2 + \omega^2}.
\end{equation}
From Eq.~\ref{eq:overlap} we have
\begin{equation}
    \left | {\expval{\eta_k|\hat{N}|\Omega}} \right |^2 = \left | {\frac{\Delta_k}{E_k}} \right |^2,
\end{equation}
which yields
\begin{align}
R_{\rm QPP}(k) &= \frac{2\pi}{\hbar^2}\frac{\Delta_k^2}{E_k^2}\frac{\delta\mu^2}{2\pi}\frac{\Gamma}{4\Gamma^2 + \omega^2} \nonumber\\
&= \frac{\delta\mu^2\Gamma \Delta_k^2}{4E_k^2(E_k^2 + \Gamma^2\hbar^2)}~.
\end{align}
Integrating over all wavevectors $k$\footnote{The integral is actually restricted to the first Brillouin zone, but since the integrand is non-negligible only near the Fermi surface, extending the integration domain to all $k$ introduces negligible error.} and approximating $\Delta_k \approx \Delta_{k_{\rm F}} = \Delta$ yields
\begin{align}
\label{eq:fermi_gold_rule_result}
R_{\rm QPP} &= \frac{\mathscr{L}}{2\pi}\int_{-\infty}^{\infty}\frac{\delta\mu^2\Gamma \Delta_k^2 dk}{4E_k^2(E_k^2 + \Gamma^2\hbar^2)} \nonumber \\
&\approx \frac{\mathscr{L}\delta\mu^2\Gamma{\Delta}^2}{2\pi}\int_{-\infty}^{\infty}\frac{dk}{4E_k^2(E_k^2 + \Gamma^2\hbar^2)} \nonumber \\
&= \frac{\mathscr{L}\delta\mu^2}{16\hbar v_{\rm F}\Delta}\Gamma(1+\hbar^2\Gamma^2/\Delta^2)^{-3/2}.
\end{align}
Therefore, we get
\begin{equation}
\mathscr{F} = (1+\hbar^2\Gamma^2/\Delta^2)^{-3/2}.
\end{equation}
In the limits of slow and fast switching rates $\Gamma$, we get
\begin{equation}
R_{\rm QPP} = \left\{ \begin{array}{lcl}
\frac{\mathscr{L}\delta\mu^2}{16\hbar v_{\rm F}\Delta}\Gamma & \text{for} & \Gamma \ll \frac{\Delta}{\hbar} \\
\frac{\mathscr{L}\delta\mu^2\Delta^2}{16\hbar^4 v_{\rm F}}\frac{1}{\Gamma^2} & \text{for} & \Gamma 
\gg \frac{\Delta}{\hbar}
\end{array}\right. .
\end{equation}
The limit for small $\Gamma$ agrees exactly with the method explained in the main text.
We can also calculate the frequency $\Gamma_{\rm max}$ at which $R_{\rm QPP}$ gets maximized.
Since $R_{\rm QPP} \propto \mathscr{F}\Gamma$, we can obtain $\Gamma_{\rm max}$ by setting
\begin{equation}
\left.\frac{d (\mathscr{F}\Gamma)}{d\Gamma}\right\vert_{\Gamma_{\rm max}} = 0.
\end{equation}
This yields
\begin{equation}
\Gamma_{\rm max} = \frac{\Delta}{\sqrt{2}\hbar} \approx \frac{4.44\Delta}{h}.
\end{equation}
This value of $\Gamma_{\rm max}$ agrees reasonably well with the numerics, 
with slight discrepancy likely to be arising from finite size effects in the simulation.
Of course, the width of the plateau is proportional to $\Delta/\hbar$.
The value of $\mathscr{F}$ at $\Gamma_{\rm max}$ is 
\begin{equation}
\mathscr{F} = (1+\hbar^2\Gamma_{\rm max}^2/\Delta^2)^{-3/2} = \frac{2\sqrt{2}}{3\sqrt{3}} \approx 0.544.
\end{equation}
We therefore get 
\begin{equation}
\mathscr{F}\Gamma\vert_{\rm max} = \frac{\Delta}{\sqrt{2}\hbar}\frac{2\sqrt{2}}{3\sqrt{3}} = 
\frac{2\Delta}{3\sqrt{3}\hbar} = \frac{2.418\Delta}{h}.
\end{equation}
This is very close to the value $2.8\Delta/h$ estimated numerically.
Finally, this yields
\begin{equation}
R_{\rm QPP,max} \approx \frac{\mathscr{L}\delta\mu^2}{24\sqrt{3}\hbar^2 v_{\rm F}}.
\end{equation}

Using $S_0 = \delta\mu^2/8$, we obtain 
\begin{equation}
    R_{\rm QPP, max} = \frac{\mathscr{L}S_0}{3\sqrt{3}\hbar^2 v_{\rm F}}~,
\label{eq:quasiparticle_excitation_rate_with_S_0}
\end{equation}
which agrees well with the numerically determined value of 
$R_{\rm QPP,max} \approx{0.7\mathscr{L}S_0}/{\pi\hbar^2 v_{\rm F}}$ reported in the main text.

\subsection{Quasiparticle excitation rate due to an ensemble of TLFs 
using Fermi's golden rule} \label{supp_sec:fermi_ensemble_TLF}
In this section, we calculate the rate of quasiparticle pair excitation resulting from an ensemble of
TLFs giving rise to 1/f noise using Fermi's golden rule. The estimate we get using this method
is of the same order of magnitude 
as the one obtained by considering the effects of individual TLFs in the limit of large TLF effective temperature.

{\color{black}
Our starting point is the expression for excitation rate 
$R_{\rm QPP}(k)$ given by Eq.~\eqref{eq:fermigoldenrule},
where now $S(\omega) = S_0/\omega$ is the spectral 
density of 1/f noise.}
From Eq.~\ref{eq:overlap}, we have
\begin{equation}
    \left |\bra{\eta_k}\hat{N}\ket{\Omega}\right |^2 = \left |{\frac{\Delta_k}{E_k}}\right |^2,
\end{equation}
which yields
\begin{equation}
R_{\rm QPP}(k) = \frac{\pi S_0 \abs{\Delta_k}^2}{\hbar E_k^3}.
\end{equation}
Now the total rate $R_{\rm QPP}$ at which quasiparticle pairs are generated is obtained by summing over the $k$ in the first Brillouin zone.  Again, extending the integration over the first Brillouin zone to all $k$ introduces negligible error, so $R_{\rm QPP}$ is given by
\begin{align}
R_{\rm QPP} =  \frac{\mathscr{L}}{2\pi} \int_{-\infty}^{\infty}R_{\rm QPP}(k) dk~.
\end{align}
As in the main text, we linearize the energy dispersion about the Fermi level, $\epsilon_k \approx \hbar v_{\rm F} \abs{k} -\mu$, 
and note that for large $v_{\rm F}$ the integral is dominated by the region when $k \approx k_{\rm F} = \mu/\hbar v_{\rm F}$. Therefore,
\begin{align}
R_{\rm QPP} &\approx \frac{\mathscr{L}}{2\pi} \int_{-\infty}^{\infty}R_k dk \nonumber\\
&= \frac{\mathscr{L}S_0}{2\hbar} \int_{-\infty}^{\infty}\frac{ \abs{\Delta_k}^2}{\sqrt{(\hbar v_{\rm F}k -\mu)^2 + \abs{\Delta_k}^2}^{3/2}}dk
\nonumber\\
&= \frac{\mathscr{L}S_0}{2\hbar^2 v_{\rm F}} \int_{-\infty}^{\infty}\frac{ 1}{\sqrt{1+z^2}^{3/2}}dz
\nonumber\\
&=\frac{\mathscr{L}S_0}{\hbar^2 v_{\rm F}}.
\end{align}
This result is the same order of magnitude  as is obtained by considering only a single TLF, which
is consistent with the arguments in Sec.~\ref{sec:Gamma_max_calculations} that including the effects of all the TLFs increases the quasiparticle 
excitation rate by a numerical factor of order unity over that of one TLF with a transition rate that maximizes the quasiparticle pair excitation rate.

\section{Numerical methods}

In this section of the Supplementary Materials we provide further details of our numerical calculations of the quasiparticle pair excitation rates from a single TLF (Sec.~\ref{supp_sec:numerics_details}). We then provide a derivation of the parameters used in our numerical calculations based on the experimental parameters reported in Refs.~\cite{Aghaee_2023_PRB_107_23,aghaee_2025_nature_638_55} (Sec.~\ref{supp_sec:nanowire_properties}). 

\subsection{Model used for simulations} \label{supp_sec:numerics_details}
Our numerical calculations have been performed using the model of a Kitaev chain~\cite{Kitaev_2001_Physics_uspekhi_44}, with
the 1/f noise incorporated via a chemical potential $\mu(t)$ that fluctuates in time.
This model has been studied previously using low-frequency noise with Gaussian time correlations~\cite{Mishmash_2020_PRB_101}.
The choice to allow $\mu$ to vary in time but be constant in space is reasonable because each InAs nanowire is immediately adjacent to superconducting Al.
The plasma frequency of Al is $\sim 2\times 10^{16}$\;\unit{\Hz}, so screening by the Al occurs very quickly on time scale corresponding to the induced gap in the InAs ($\sim 2.5\times 10^9$\;\unit{\Hz}), resulting in a uniform chemical potential shift of the entire nanowire. 
We reproduce the Hamiltonian here for convenience:
\begin{equation}
   \hat{H}_{\rm K} = \sum_{i=1}^N \left [ -\mu(t) c_i^\dagger c_i - \frac{1}{2} \left (w c_i^\dagger c_{i+1} + \Delta c_i c_{i+1} + \rm{H.c.} \right )
   \right ]~.
   \label{eq:Kitaev_supplemental}
\end{equation}
Here, $c_i^\dagger$ and $c_i$ are operators that create and annihilate spinless fermions at site $i$ respectively,  {\color{black} $\mu(t)$ is the chemical potential of the Kitaev chain, where the time-dependence has been made explicit}, 
$w$ is the nearest-neighbor hopping strength of the fermions,
$\Delta$ is the superconducting gap,
and H.c.\ denotes the Hermitian conjugate. 
{\color{black} Based on the device parameters reported in Ref.~\cite{aghaee_2025_nature_638_55}},
we investigate superconducting gaps $\Delta=\Delta_{k_{\rm F}}$ ranging from $10$ to $\sim 200~\mu$eV ($\Delta/h$ ranging from $\sim 2.42-48$~GHz) and a nanowire
length $\mathscr{L}=$ 3\;\unit{\um}.
A single TLF with switching rate $\Gamma$ 
is modeled by having the chemical potential repeatedly switch between values 
$\mu_1 = 0$ and $\mu_2 = 0.5657$\;\unit{\micro\eV}.

We perform the simulations efficiently
using the covariance matrix framework presented in Ref.~\cite{Surace_2022_SciPost_Lec}. 
The methods used for the numerical calculations, which are described in detail in Ref.~\cite{goffage2025leakage}, are similar to those used in Refs.~\cite{Mishmash_2020_PRB_101,Surace_2022_SciPost_Lec}.
The numerical simulations are performed on the Kitaev Hamiltonian given in Eq.~\eqref{eq:Kitaev_supplemental}, which does not include explicit dissipative mechanisms such as electron-phonon interactions or electron-electron interactions.  Including these mechanisms into the simulations is not feasible because the simulation methods rely on the quadratic nature of the Kitaev Hamiltonian (otherwise an exponentially large Hilbert space would be needed), as discussed in Ref.~\cite{Surace_2022_SciPost_Lec}. 

The calculations shown in Fig.~2 of the main text show that the rate at which quasiparticles are excited in a Kitaev chain with a single two-level fluctuator (TLF) that switches the chemical potential $\mu$ between two different values with switching rate $\Gamma$ increases with $\Gamma$ for $\Gamma$ up to a frequency $\Gamma_{\rm max}$ that is close to $4\Delta/h$.
We calculate the rate of quasiparticle pair formation in the presence of a single TLF as opposed to a population of TLFs with different frequencies.
This choice provides a lower bound to the quasiparticle pair production rate and also provides a reasonable estimate of the actual rate because the quasiparticle pair production is dominated by a relatively small number of TLFs with $\Gamma$'s in the region of the broad maximum.
{The calculations for a single TLF} simplify substantially:
(1) between the jumps, the Hamiltonian is time-independent.
Therefore, the operator $e^{iH_{\rm BdG}t}$ that governs the time evolution can be computed efficiently by diagonalizing the Hamiltonian $H_{\rm BdG}$ and then time-evolving the correlation matrix through a simple matrix product 
(see \cite{Surace_2022_SciPost_Lec} for further details); and (2) because $\mu$ takes on two values, only two diagonalizations need to be performed.
Because of this simplicity, it is straightforward to perform calculations on system sizes comparable to or larger than those being used experimentally. Note that $H_{\rm BdG}$ is the Bogoliubov-de Gennes Hamiltonian which can be derived in a straightforward manner from $\hat{H}_{K}$ as explained in detail in \cite{Surace_2022_SciPost_Lec}. We also remark that our simulations are carried out for the Kitaev-tetron Hamiltonian which consists of two uncoupled Kitaev chains (in the same manner as Refs.~\cite{Mishmash_2020_PRB_101, goffage2025leakage}); however, we report the quasiparticle excitation probability for a single chain. 



In the succeeding section~\ref{supp_sec:nanowire_properties} we obtain numerical values of Kitaev chain parameters using the information reported in Refs.~\cite{Aghaee_2023_PRB_107_23} and~\cite{aghaee_2025_nature_638_55}.

\section{Determination of Kitaev chain parameters used in numerical calculations} \label{supp_sec:nanowire_properties}
In Fig.~2 of the main text we present the probability of exciting at least one quasiparticle pair ($R_{\rm QPP}$) for a Kitaev chain with parameters obtained from the nanowire properties of the Microsoft Azure Quantum experiments as reported in  \cite{Aghaee_2023_PRB_107_23} and \cite{aghaee_2025_nature_638_55}. 
\textcolor{black}{Here we show the determination of Kitaev chain parameters from these experimental nanowire properties. The results are summarized in Table~1.} 

The parameters that need to be determined are $\Delta$ (superconducting gap amplitude), $\delta\mu$ (noise amplitude) $w$ (hopping amplitude), $a$ (lattice constant), and $N$ (chain length). 
We examine a range of superconducting gap amplitudes $\Delta$ ranging up to 200~$\mu$eV.
The magnitude of the chemical potential of the nanowire from fluctuation from a TLF is determined from the charge noise amplitude along with the capacitance of the nanowire.
In Ref.~\cite{aghaee_2025_nature_638_55} the coefficient $S_0$ of the chemical potential fluctuations (noise spectral power density = $S_0/\omega$) at a dot with capacitance 5 times smaller than the nanowire was measured to be $(1~\mu{\rm eV})^2$.
Supplementary Sec.~\ref{supp_sec:delta_mu_derivation} shows that this value of $S_0$ corresponds to a value of $\delta\mu$ for the nanowire of
$\delta\mu = \sqrt{8S_0}/(C_w/C_d)$, where $C_w$ is the capacitance of the nanowire and $C_d$ is the capacitance of the dot at which $\delta\mu$ was measured.
This yields a value of $\delta\mu$ for the nanowire of $\delta\mu = 0.566~\mu{\rm eV}.$
Our numerical calculations are done for a single TLF, so at any given time $t$,  $\mu(t) \in \{\mu_1,~\mu_2\}$ where $\delta \mu = \mu_2 - 
\mu_1$ and $\mu_1 = 0$\;\unit{\micro\eV}.

The chain length $N$ is the ratio of the nanowire length $\mathscr{L}$ and the lattice site constant $a$ in the Kitaev chain model.
We set $a$ to be $\pi/(2 k_F)$, where $k_F$ is the Fermi wavevector.
With this choice, when $\mu=0$ the Fermi wavevector $k_F$ satisfies $k_F=\pi/2a$.

We use a value of the Fermi wavevector of $k_{\rm F}^{-1} = 40$\;\unit{\nm},  which is in the range given Ref.~\cite{Aghaee_2023_PRB_107_23},
and find
\begin{align}
    a = \frac{\pi}{2 k_{\rm F}} = \frac{\pi}{2}\times 40\times10^{-9}\;\unit{\m} = 63\;\unit{\nm} . 
\end{align}
For a nanowire of length $10$\;\unit{\um}, this in turn gives the number of sites in the Kitaev chain of
\begin{align}
    N = \left\lfloor \frac{\mathscr{L}}{a} \right\rfloor = \left\lfloor \frac{10\;\unit{\um}}{0.062832\;\unit{\um}}\right\rfloor = 159.
\end{align}
Similarly we obtain values of $N$ for $\mathscr{L} = 3$\;\unit{\um} and $5$\;\unit{\um} as summarized in Table~1. 

\begin{figure}[!ht]
    \centering
    \includegraphics[width=0.5\linewidth]{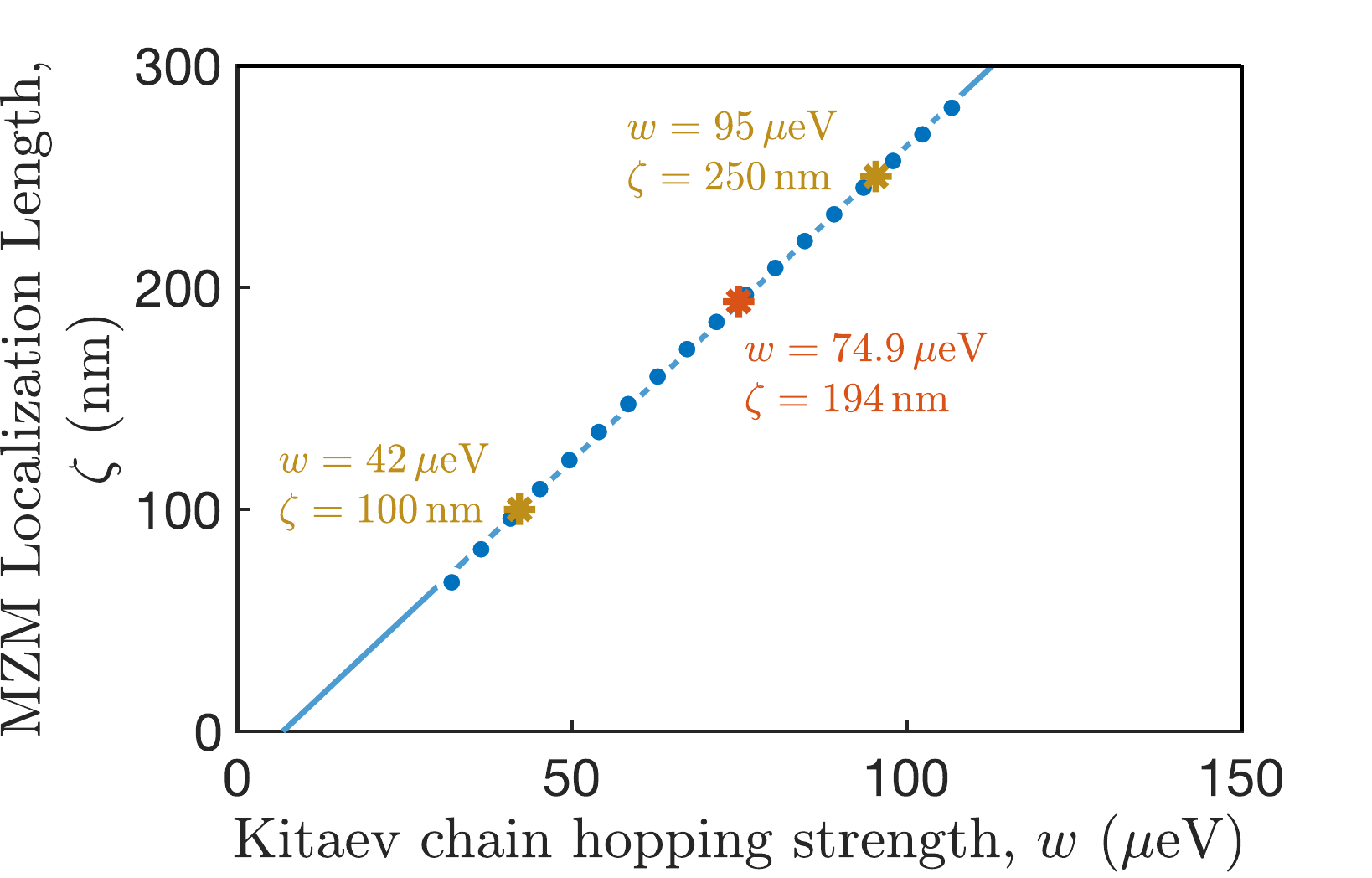}
\caption{\small {\bf Majorana zero mode localization (MZM) length $\zeta$ versus hopping $w$ in the Kitaev chain model.} This calculation was performed to determine the value of $w$ based on the value of $\zeta$ reported in experiment~\cite{aghaee_2025_nature_638_55}.  Here, based on experimental numbers reported in Ref.~\cite{aghaee_2025_nature_638_55}, we have taken ${\mathscr{L}} = 10\;\unit{\um},~\Delta = 23.48\;\unit{\micro\eV}$,  $\mu = 0\;\unit{\micro\eV}$, and lattice constant 
$a = 63$\;\unit{\nm}, so that the number of lattice sites in the Kitaev chain is $N = 159$. Numerically calculated localization lengths are denoted by dots and asterisks, and the linear fit is denoted by the solid blue line. The yellow asterisks show $w$ and $\zeta$ at the upper and lower bounds of $\zeta$ as reported in Ref.~\cite{Aghaee_2023_PRB_107_23}. The orange asterisk shows the $w$ and $\zeta$ as used in the numerical results reported in the main text. The MZM localization length $\zeta$ is determined by fitting $Ae^{-x_j/\zeta}$ to the MZM wavefunctions where $x_j = aj$ (and $j = 1,2, ... N$ is the index of the $j$'th lattice site).} 
\label{fig:supp_local_length}
\end{figure}

We find appropriate values for the hopping parameter $w$ of the Kitaev model by ensuring the Majorana zero mode (MZM) localization length $\zeta$ is significantly smaller than the nanowire length (based on the discussion in Ref.~\cite{Aghaee_2023_PRB_107_23}, we set it to $\sim$0.25~$\mu$m). First note that deep in the topological regime, {\color{black}$\zeta \approx \xi$} where $\xi$ is the superconducting coherence length~\cite{Alicea_2012_IOP_75}. 
\textcolor{black}{Fig.~\ref{fig:supp_local_length} presents numerically calculated MZM localization lengths $\zeta$ versus $w$ for a Kitaev chain with $N = 159,~{\Delta} = 23.48~\mu$eV, lattice constant $a=63$~nm, and ${\mu} =0$. 
For this value of $\Delta$, we obtain 
$\zeta \approx 194$~nm for
a value of $w\sim 74.9~\mu$eV.  
(The rest of this section uses this value).
At $\mu=0$, the value of $w$ at which $\zeta$ has a given value is proportional to $\Delta$, so we can write
$w=A\Delta$, with $A=(dw/d\Delta) = (74.9/23.48)=3.19$. 
Since $v_F = w a/\hbar$~\cite{Alicea_2012_IOP_75}, where the lattice constant $a=\pi/(2 k_F)$, we find 
$v_F=A\Delta a/\hbar=3.19\Delta a/(6.582 \times 10^{-10}~\mu$eV-s).
Using $a=63$~nm, we find, for $\Delta$ given in microvolts, $v_F = (3.19)\Delta(63~{\rm nm})/\hbar=30500\Delta$~cm/s, where again, $\Delta$ is in $\mu$eV.}




\section{Characterization of the dependence of the quasiparticle pair excitation rate $R_{\rm QPP}$ on the TLF effective temperature $T_{\rm eff}$.}
\label{supp_sec:thermal}
This subsection presents details of our calculations of the rate of quasiparticle pair excitation as a function of the TLF effective temperature.
It demonstrates that the number of quasiparticle pairs excited by 1/f noise is likely to be orders of magnitude larger than would be indicated by just considering the value of $\exp(-2\Delta/k_B T)$, where $\Delta$ is the magnitude of the topological gap and $T$ is the electron temperature.
The main reason that the quasiparticle pair excitation rate is larger than naive expectations is that 
microwave fields for qubit readout as well as large electric fields from the pulsed gate voltages used for qubit manipulation~\cite{Muller_2019}
couple strongly to TLFs and drive them out of equilibrium, so that $T_{\rm eff}$, the effective temperature of the TLFs, is likely to be significantly greater than the electron temperature.
Indeed, there is quite direct evidence from experiments that in silicon quantum dot qubits $T_{\rm eff}>0.3$~K at electron temperatures down to $0.14$~K~\cite{Huang:2024p772}.
The methods of qubit manipulation and measurements of silicon spin qubits are very similar to those of Microsoft's implementation of Majorana qubit, so the Si experimental measurements of large values of $T_{\rm eff}$ are highly relevant to the hybrid superconductor-semiconductor system.
As shown in Fig.~\ref{fig:F2} in the main text, if the TLF effective temperature is similar to that measured in Si qubits, substantial quasiparticle pair excitation rates are found at the values of the topological gap reported in Ref.~\cite{aghaee_2025_nature_638_55}.

This section also discusses a second effect that leads to larger quasiparticle pair excitation rates that operates if the TLFs are thermally excited over an energy barrier (as originally proposed by Dutta and Horn~\cite{Dutta:1981p497}).
If the TLFs are thermally activated,
 the large attempt frequencies and energy barriers of TLFs in typical semiconductors mean that TLFs at GHz frequencies are activated over energy barriers that are much larger than the temperature, and therefore the energy available to excite quasiparticles is significantly larger than the effective temperature of the TLFs.
If both factors operate, then the superconducting gaps would need to be an order of magnitude larger than in current experiments to significantly suppress the rate of quasiparticle pair excitation.
The calculations for thermally activated TLFs also explicitly justify the assertion that the time scale of each TLF transition is very short, and we argue that a similar (though somewhat less rigorous) method can be made to justify the assumption for TLFs whose transitions are made via quantum tunneling.

\subsection{
Evidence that the TLF effective temperature $T_{\rm eff}$ is significantly larger than the electron temperature in quantum dot qubit systems.
}
\label{subsec:TLF_effective_temperature_data}
Microsoft's Majorana devices use a semiconductor-superconductor hybrid architecture, where the qubit measurements and operations are performed by measurements made using quantum dots to which voltages are applied to adjust their chemical potentials by substantially more than $kT$, where $T$ is the electron temperature.
Because many TLFs have electric dipole moments, these dynamic voltage changes have the potential to heat the TLFs so that the TLF effective temperature $T_{\rm eff}$ is significantly higher than the electron temperature $T$.
Recent measurements of silicon quantum dot qubits~\cite{Huang:2024p772} provide strong evidence that this indeed occurs.

\begin{figure}[!t]
  \centering
\includegraphics[width=0.97\linewidth]{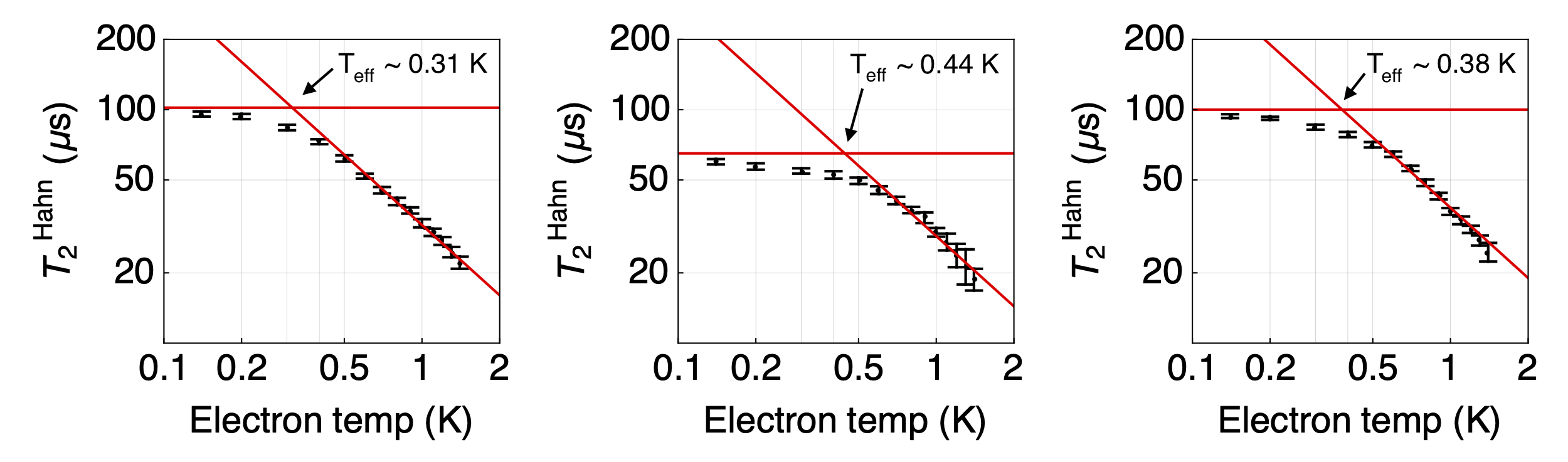}
\caption{Experimental data of the dependence on electron temperature of the $T_2^{\rm Hahn}$ coherence time for three different configurations of a single silicon spin qubit, from Refs.~\cite{Huang:2024p772,https://doi.org/10.5281/zenodo.10452860}.The relevant data points from Fig.~3b of Ref.~\cite{Huang:2024p772} are replotted here with lines drawn to emphasize that this decoherence time, which is determined by charge noise from TLFs~\cite{struck2020low}, depends on the electron temperature $T$ as $1/T$ at relatively high $T$, but then saturates as the temperature is lowered at an electron temperature of $0.3$~K or greater.
We estimate the TLF effective temperature $T_{\rm eff}$ as the electron temperature at which the $T_2^{\rm Hahn}$ observed at asymptotically low temperature intersects the $1/T$ extrapolation.
For all three data sets, this method yields an effective temperature $T_{\rm eff}$ of $0.3$~K or higher.
}
\label{supp_fig:Huang_data_hot_qubits}
\end{figure}

Figure~\ref{supp_fig:Huang_data_hot_qubits} shows experimental measurements of $T_2^{\rm Hahn}$ of a silicon spin qubit, which is believed to be determined by charge noise from TLF fluctuations~\cite{struck2020low}.
Within the standard TLF model~\cite{Dutta:1981p497}, as $T_{\rm eff}$ is lowered, this decoherence rate decreases as $1/T_{\rm eff}$, because the standard TLF distribution has a uniform distribution of asymmetries $\varepsilon$, and the range of accessible $\varepsilon$ is proportional $T_{\rm eff}$.
The experimental data for $T_2^{\rm Hahn}$ follow a $1/T$ dependence at higher electron temperatures $T$, but this
dependence saturates at low $T$.
This saturation is evidence that the TLFs are not in thermal equilibrium with the electrons at low electron temperatures and that $T_{\rm eff}$ is higher than $T$.
We estimate $T_{\rm eff}$ by extrapolating the $1/T$ behavior at higher electron temperatures and then finding the electron temperature at which the low-temperature value of the dephasing rate is achieved.
As shown in Fig.~\ref{supp_fig:Huang_data_hot_qubits}, this extrapolation procedure yields $T_{\rm eff}$ of the TLFs for this silicon spin qubit system of $\sim 0.31$~K, $\sim 0.44$~K, and $\sim 0.38$~K.

It is reasonable to assume that the TLFs in Microsoft's Majorana system also have a $T_{\rm eff}$ that is similar to that measured in the silicon system.
Therefore, we compare
these effective temperatures to the values of $2\Delta$, where $\Delta$ is the topological gap, reported in Ref.~\cite{aghaee_2025_nature_638_55}.
The 90$^{\rm th}$ percentile of the values of $\Delta$ is 23.8~$\mu$eV.
For an effective temperature of $T_{\rm eff}=0.3$~K, $\exp(-2\Delta/k_B T_{\rm eff})=0.159,$
which is a suppression by less than
one order of magnitude.

\subsection{Effects of large TLF attempt frequencies on the dependence on the TLF effective temperature of the quasiparticle pair excitation rate.}
\label{supplemental_subsec:attempt_frequency}
This subsection investigates another reason why the thermal suppression of the excitation of quasiparticle pairs by 1/f noise is significantly less effective than would be guessed by looking at the value of $\exp(-2\Delta/k_B T).$
In the previous subsection we showed that the effective temperature of the TLFs can be significantly greater than the electron temperature.
In this subsection we show that significant creation of quasiparticle pairs can be induced by 1/f noise even if the superconducting gap energy is an order of magnitude larger than $k_B T_{\rm eff}$, where $T_{\rm eff}$ is the temperature of the TLF ensemble.  
This apparently paradoxical situation occurs because 
TLF attempt frequencies are very high, of order the Debye frequency~\cite{Vineyard:1957p121}, which in InAs is about 5.8~THz.
The large attempt frequency means that thermally excited TLF transitions at frequency $k_B T/h$ occur only when the TLF is excited over the large energy barrier $E_b$, and the $E_b$ is then available to excite quasiparticle pairs over the superconducting gap.
We take this availability of energy into account and find that substantial quasiparticle pair excitation will occur for the experimentally reported gaps in Ref.~\cite{aghaee_2025_nature_638_55} even in thermal equilibrium at the electron temperature of 50~mK reported in the absence of leads and with no pulsed voltages applied~\cite{aghaee_2025_nature_638_55}.

\begin{figure}
\vskip -2.5cm
    \centering
    \includegraphics[width=1\linewidth]{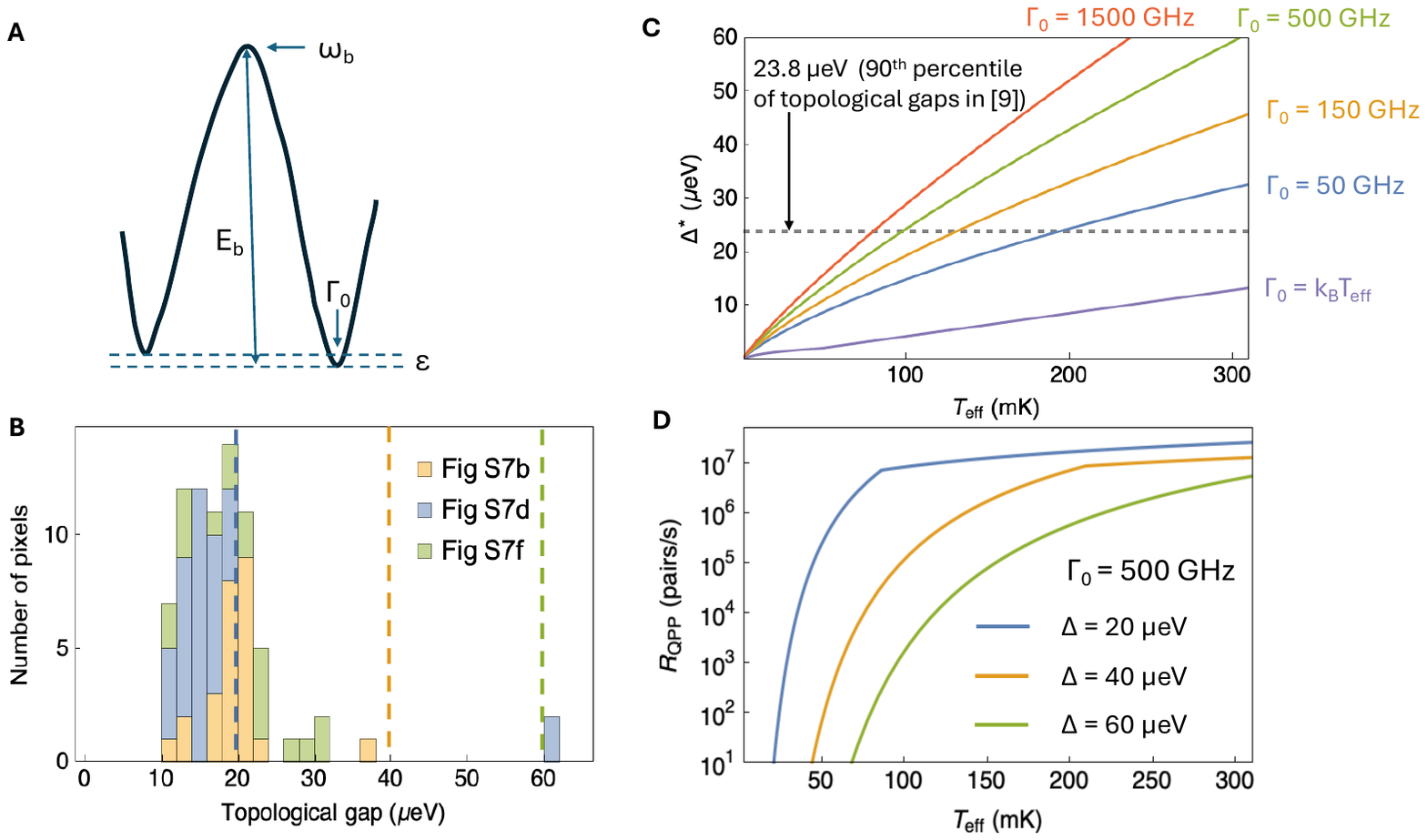}
    \vskip -1.5 cm
    \caption{Effects of a large TLF attempt frequency on quasiparticle pair excitation rate $R_{\rm QPP}$ for classically activated TLFs.  A: Schematic of a TLF, with the attempt frequency $\Gamma_0$, the barrier height $E_b$, the barrier frequency arising from the curvature at the top of the barrier $\omega_b$, and the TLF asymmetry $\varepsilon$ indicated.  
    {\bf B}: Histogram of measured values of the topological gap $\Delta$, digitized from Fig.~S7 of Ref.~\cite{aghaee_2025_nature_638_55}. 
    The colors denote measurements of three different device configurations.
    The median gap $\Delta$ is 16.8~$\mu$eV and the 90$^{\rm th}$ percentile is 23.8~$\mu$eV.
    The dashed vertical lines indicate the values of $\Delta$ used for the curves shown in D.
    {\bf C}: Plot of $\Delta^*$, the value of $\Delta$ that satisfies $k_BT_{\rm eff}\ln(h\Gamma_0/k_BT)=2\Delta$, versus the TLF effective temperature $T_{\rm eff}$. Increasing the attempt frequency $\Gamma_0$ causes $\Delta^*$ to increase because each transition of a TLF with a large $\Gamma_0$ occurs rarely, when the TLS thermal bath provides the energy to surmount the barrier $E_b$, which is then available to excite the quasiparticle pairs over the energy gap 2$\Delta$.  
    {\bf D}: Rate of quasiparticle pair excitation $R_{\rm QPP}$ versus $T_{\rm eff}$ for three values of $\Delta$, for a 3~$\mu$m InAs nanowire, with materials parameters appropriate to the experiments reported in Ref.~\cite{aghaee_2025_nature_638_55} (see Table~1).  The 
    rates are plotted versus $T_{\rm eff}$ with $\Gamma_0$~=~500~GHz.  With this value of $\Gamma_0$, which is well below the InAs Debye frequency of $\sim5.8$~THz, $R_{\rm QPP}$ is substantial for $20~\mu$eV gaps even for $T_{\rm eff}$ significantly below 50~mK.}
    \label{supp_fig:effective_temperature_with_barrier}
\end{figure}

We adopt the model of Dutta and Horn~\cite{Dutta:1981p497}, where the two-level fluctuators are thermally activated over energy barriers.
We consider the dynamics of a single TLF, as illustrated in Fig.~\ref{supp_fig:effective_temperature_with_barrier}{\bf A}.

The relevant time scales for a single TLF making transitions that are thermally excited over an energy barrier are the first passage time, which is the typical time between changes of the TLF from one state to the other~\cite{Hanggi:1990p251}, and the
transition path time, which is the typical time of the set of paths that start in one state and end at the other without revisiting the starting position~\cite{Kim:2015p224108}.
We identify the transition rate $\Gamma$ of a TLF in the main text with the inverse of the first passage time, which has the well-established activated Kramers form~\cite{Hanggi:1990p251}:
\begin{equation}
\Gamma \approx \Gamma_0 \exp(-E_b/k_B T_{\rm eff})~,
\label{eq:Gamma_in_thermal_equilibrium}
\end{equation}
where $\Gamma_0$ is the attempt frequency, $E_b$ is the energy barrier height, $k_B$ is Boltzmann's constant, and $T_{\rm eff}$ is the TLF effective temperature.
The sharpness of the individual transitions is determined by the transition path time $\tau_{TP}$, which depends only logarithmically on $E_b$~\cite{Kim:2015p224108}:
\begin{equation}
\tau_{TP} \approx \frac{1}{\omega_b}\ln \left (\frac{2 e^\gamma E_b}{k_B T} \right )~,
\label{eq:transition_path_time}
\end{equation}
where $\omega_b$ is a bare frequency describing the top of the potential barrier and $\gamma$ is Euler's constant ($\approx 0.577$).

First we discuss the transition path time $\tau_{TP}$.
The dependence of $\tau_{TP}$ on $E_b/k_B T$ is only logarithmic, so $\tau_{TP}$ is of order $\frac{1}{\omega_b}$ even when $\Gamma$ is many orders of magnitude smaller than $\Gamma_0$.
Therefore, if the microscopic frequencies $\Gamma_0$ and $\omega_b$ are both much larger than $\Delta$, then when $2\Delta/k_BT \gg 1$, the inverse of the transition path time remains of order the microscopic frequency even when $\Gamma$ is very low, a result that demonstrates that the time scale of the TLF transitions is extremely short and therefore the resulting time dependence of the chemical potential has Fourier components that extend out to frequencies greater than $2\Delta/k_BT$.

We also note that a similar but less rigorous argument can be made that for a TLF that is tunneling incoherently between its states, the quantum transitions occur on a time scale that is much less than its transition rate when the tunneling rate is much less than the attempt frequency.
While the tunneling time is not as well defined as the classical transition path time, Ref.~\cite{Buttiker:1982p1739} presents a sensible approach to calculating it by considering the effects of modulating the barrier in time.
This approach yields a time $\tau$ (denoted in~\cite{Buttiker:1982p1739} as the interaction time) that actually decreases with increasing barrier height:
\begin{equation}
\tau =\int_{x_1}^{x_2} dx 
 \left [ \frac{m}{2 \left ( V_0(x)-E \right ) } \right ]^{1/2}~,
\end{equation}
where $V_0(x)$ is the potential, $E$ is the energy, $x_1$ and $x_2$ are turning points, and $m$ is the mass.
This time actually gets shorter as the barrier height increases, again demonstrating that the individual transitions occur much faster than the time between successive tunneling events.

Now we calculate the rate of excitation of quasiparticle pairs out of the condensate in a system at thermal equilibrium at temperature $T$.
This rate can be nonnegligible even when $k_B T \ll \Delta$ because the attempt frequency $\Gamma_0$ is vastly larger than the transition rate $\Gamma$; this follows because attempt frequencies are typically of order the Debye temperature~\cite{Vineyard:1957p121}, so the fluctuations of the thermal bath that cause TLF transitions are extremely atypical.  
Using Eq.~\ref{eq:Gamma_in_thermal_equilibrium}, we find the barrier height $E_b$ that leads to a transition rate $\Gamma$ is:
\begin{equation}
E_b = k_B T \ln \left (\Gamma_0/\Gamma \right )~.
\label{eq:Gamma_dependence_on_E_B}
\end{equation}
With an attempt frequency and a frequency of the barrier top of 4~THz (corresponding to $\sim$200~K, which is somewhat less than the Debye temperature~\cite{Vineyard:1957p121} of InAs of 280K), the TLF
energy barriers for $h\Gamma=k_BT_{\rm eff}$ at different effective temperatures are $E_b({\rm 50~mK}) \approx 35~\mu$eV (equivalent to $\approx 0.41$~K),
$E_b({\rm 100~mK})\approx 65~\mu$eV (equivalent to $\approx 0.76$~K), and
$E_b({\rm 200~mK})\approx 118~\mu$eV (equivalent to $\approx 1.37$~K).
When the energy barrier of the relevant TLF exceeds 2$\Delta$, where $\Delta$ is the energy of the topological gap, the fact that the TLF was excited over its energy barrier implies that enough energy is available to excite a quasiparticle pair.
However, if 2$\Delta$ exceeds $E_b$, then quasiparticle excitation is suppressed.
As described in the Methods, we account for these effects by adding exponential cutoffs, yielding Eq.~\ref{eq:R_QPP_with_thermal_activation}, which we reproduce here for convenience:
\begin{equation}
R_{\rm QPP} = \max_\Gamma 
\frac{\mathscr{L}}{16}\frac{(\delta\mu)^2\Gamma}{\hbar v_F\Delta}
\exp\left [-\left (\max\left (0,\frac{2\Delta}{k_B T_{\rm eff}}-\frac{\Gamma_0}{\Gamma} \right )+\frac{h\Gamma}{2\Delta}\right )\right ]~.
\end{equation}
Fig.~\ref{supp_fig:effective_temperature_with_barrier}D shows results obtained using this approach for an attempt frequency of 4~THz.
Substantial quasiparticle pair excitation rates are observed at temperatures even below 50~mK for more than 90\% of the values of the gap measured in Ref.~\cite{aghaee_2025_nature_638_55}.
Our methods for obtaining the values of the topological gap measured in Ref.~\cite{aghaee_2025_nature_638_55} are presented in the following section.

\section{Construction of the histogram of values of the topological gap measured in Ref.~\cite{aghaee_2025_nature_638_55}}
\label{supp_sec:gap_histogram}
The values of the topological gap $\Delta$ measured in Ref.~\cite{aghaee_2025_nature_638_55} were obtained by digitizing the figure panels S7b, S7d, and S7f in the supporting material of~\cite{aghaee_2025_nature_638_55} using the 
software package~\textsc{ImageJ} and calibrating the magnitudes by also digitizing the scale bars for each image.
The histograms shown in Fig.~\ref{supp_fig:effective_temperature_with_barrier}{\bf C} plot
the number of observations of measurements of $\Delta$ as a function of the magnitude of $\Delta$ using bins of size 0.2~$\mu$eV.

\section{Motion of quasiparticle pairs after excitation and implications for Majorana qubit parity lifetime}
\label{supp_sec:diffusion}
In this section we discuss the motion of  quasiparticles in the pairs excited by 1/f noise and the implications of this motion for tetron qubit coherence.
In particular, this section shows that excitation of quasiparticle pairs does not cause measurements of the Majorana parity to decay if the time it takes for quasiparticles to be absorbed by the Majorana modes is much less than both the rate of creation of quasiparticle pairs and the time resolution of the measurement.\
If both quasiparticles are absorbed in a time that is less than both the time resolution of the measurements and the mean time between quasiparticle excitation events, then measurements of the parity will not detect any change.
The time resolution of the parity measurements is about a microsecond~\cite{aghaee_2025_nature_638_55,Aghaee.arxiv.2507.08795}, and the mean time between quasiparticle pair excitation events is at least a microsecond (see, e.g., Fig.~\ref{fig:F2}), and we show here that the mean time for quasiparticles to be absorbed by the Majorana modes is much less than a microsecond.

Our calculations use the results of Karzig {\it et al.} (Ref.~\cite{karzig_2021_phys_rev_let_126}), who show that 
Majorana modes absorb quasiparticles efficiently.
(It will turn out that the time for excited quasiparticles to reach the ends of the wire is very short, so that the Majoranas will be absorbed with time to spare even if the absorption is not as efficient as predicted, but we do require that either both quasiparticles are absorbed by Majoranas or both quasiparticles decay via another mechanism.  If one quasiparticle is absorbed by a Majorana and the other decays via another mechanism, the parity will change.)
First we consider a perfectly clean nanowire in which quasiparticles move ballistically.
For this case, the two quasiparticles will travel to opposite ends of the wire with a velocity that is of order the Fermi velocity, and the time $\mathcal{T}$ to travel a distance $\mathscr{L}=3~\mu$m is $\mathcal{T}=\mathscr{L}/v_F$.
For a typical value of $v_F=10^6$~cm/s $=10^{10}\mu$m/s, $\mathcal{T}\approx $ 0.3~ns.

Now we discuss the case where the nanowires have defects, so that the quasiparticle motion is diffusive.  For this case,
the time for a quasiparticle to diffuse a distance $\mathscr{L}=3\mu$m is given by $\tau=\mathscr{L}^2/D$, where the diffusion constant $D=v\ell$ and $\ell$ is the mean free path, which we take to be the Fermi wavelength.
Using the values from Table~1, $\mathscr{L}=3\mu$m and $\lambda_F \approx 40~{\rm nm}$, and a typical value of $v_F$ of $10^6~{\rm cm/s}$, we find $\tau\approx 23~$ns.
This time is also much shorter than both the time between quasiparticle pair excitation events and the temporal resolution of the measurement.



\section{Quasiparticle poisoning for the hexon qubit} \label{supp_sec:hexon}
\begin{figure}[!ht]
    \centering
    \includegraphics[width=0.9\linewidth]{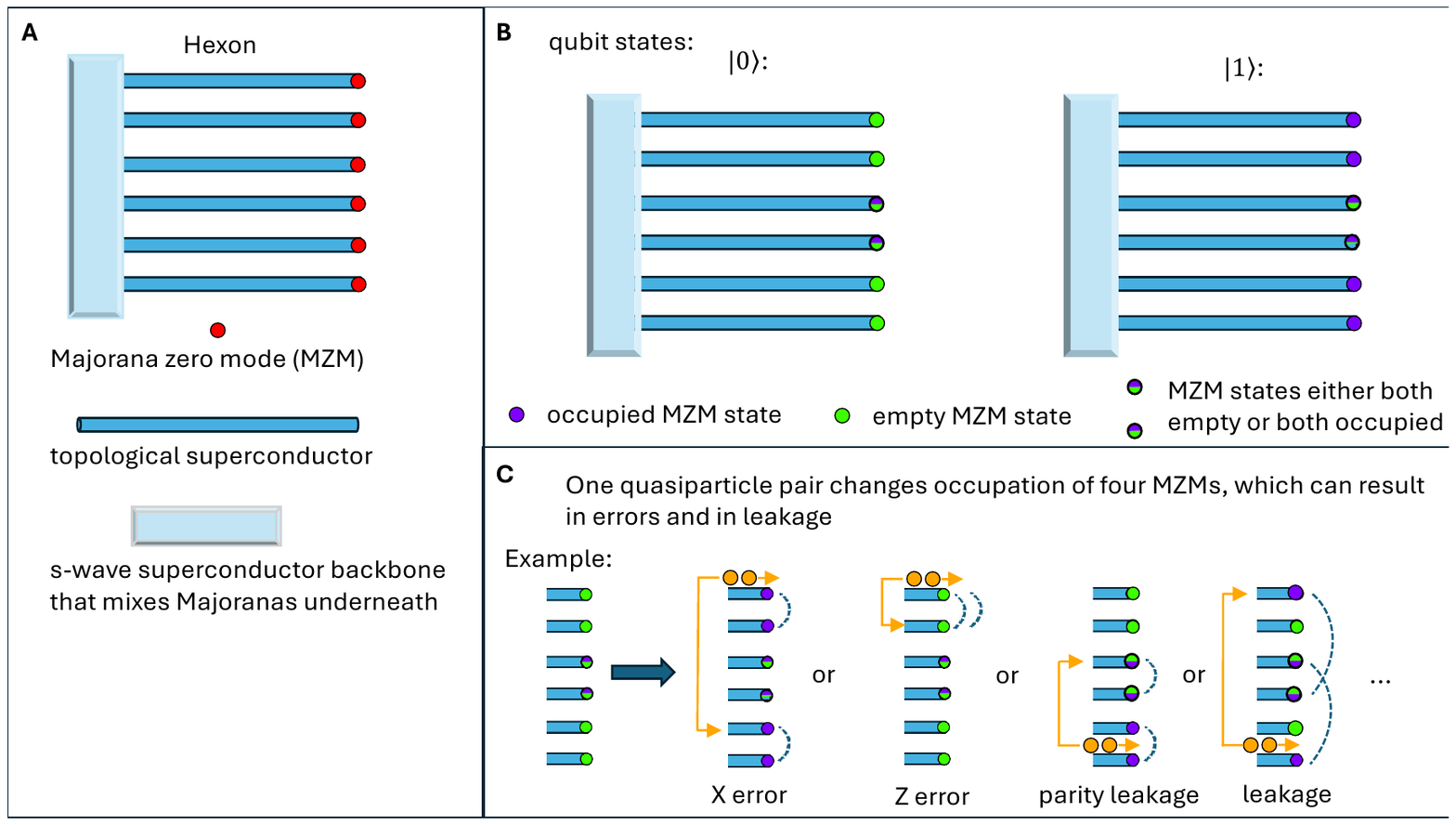}
    \vskip -0.5cm
    \caption{\small {\bf Hexon qubit and decoherence arising from quasiparticle excitations.}
  {\bf A:} Schematic of hexon qubit~\cite{Karzig_2017_Phys_Rev_B_95}, which consists of six nanowires of topological superconductor along with a  backbone of s-wave superconductor designed so that the MZM modes in the nanowires underneath exhibit strong mixing with each other.
 {\bf B:} Qubit states of a hexon qubit.  The top two and bottom two MZMs are constrained to be in an even-parity state, with the inner two MZMs are constrained to be in an odd-parity state.
{\bf C:} Excited quasiparticles in a hexon qubit can induce both qubit errors and leakage, depending on which pair of wires the MZM occupations change. There are a large number of possibilities, most of which result either in a qubit error or in leakage to a non-qubit state.
This complexity makes correcting errors arising from the excitation of quasiparticle pairs more difficult.
}
    \label{fig:hexon}
\end{figure}

Here we discuss the effects of quasiparticle pair excitation on the hexon qubit, which is a topological qubit with a different arrangement of nanowires.
The hexon qubit proposed in Ref.~\cite{Karzig_2017_Phys_Rev_B_95} and investigated in Ref.~\cite{tran2020optimizing} consists of six nanowires connected by an s-wave superconducting backbone that strongly mixes the Majorana modes of all six nanowires, as shown in Fig.~\ref{fig:hexon}.

The nanowires in the hexon that are under the s-wave superconducting backbone are engineered to be close enough that the MZMs under the backbone are all strongly mixed.
Therefore, a quasiparticle that is excited in any of the nanowires can travel to any of the other nanowires.  This additional freedom is useful when quasiparticles are absent because it increases the flexibility of measurement-based operations for qubit manipulation, but it leads to additional complexities when quasiparticle pairs have been excited.
Fig.~\ref{fig:hexon}
illustrates that this increased number of possible paths that the quasiparticles can take leads to a variety of qubit errors and leakage states.
This complexity makes error correction much more challenging than in situations with fewer possible errors.
Because the basic scaling of the quasiparticle excitation and motion is similar for the hexons and tetrons, the fundamental problem that the error rate is faster than the qubit manipulation speed will be present in the hexon architecture.

\newpage



\clearpage 





\end{document}